\documentclass[aps,pra,preprint,showpacs,floatfix,amsmath]{revtex4}

\usepackage{graphicx}
\usepackage{hyperref}
\usepackage{bm}

\begin{document}
\title{Temporal Interferometry: A Mechanism for Controlling Qubit 
        Transitions During Twisted Rapid Passage with Possible
        Application to Quantum Computing}

\date{\today}

\author{Frank Gaitan}
\email{gaitan@physics.siu.edu}
\affiliation{Department of Physics; Southern Illinois University;
              Carbondale, IL 62901-4401}

\begin{abstract}
In an adiabatic rapid passage experiment, the Bloch vector of 
a two-level system (qubit) is inverted by slowly inverting an external
field to which it is coupled, and along which it is initially aligned.
In twisted rapid passage, the external field is allowed to twist around 
its initial direction with azimuthal angle $\phi (t)$ at the same time 
that it is inverted. For polynomial twist: $\phi (t) \sim Bt^{n}$. 
We show that for $n \geq 3$, multiple avoided crossings can occur during
the inversion of the external field, and that these crossings give rise 
to strong interference effects in the qubit transition probability.
The transition probability is found to be a function of 
the twist strength $B$, which can be used to control the time-separation of 
the avoided crossings, and hence the character of the interference. 
Constructive and destructive interference are possible. The interference 
effects are a consequence of the temporal phase coherence of the wavefunction.
The ability to vary this coherence by varying the temporal separation
of the avoided crossings renders twisted rapid passage with adjustable
twist strength into a temporal interferometer through which qubit transitions
can be greatly enhanced or suppressed. Possible application of this 
interference mechanism to construction of \textit{fast fault-tolerant} 
quantum CNOT and NOT gates is discussed.
\end{abstract}


\maketitle

\section{\label{sec1}Introduction}

Adiabatic rapid passage (ARP) is a well-known procedure for inverting
the Bloch vector of a two-level system (qubit) \cite{arp}.
This is accomplished by inverting an external field $\mathbf{F}(t)$
which couples to the qubit, and along which the qubit is initially
aligned. The field inversion is done on a time-scale that is large
compared to the inverse Rabi frequency $\omega_{0}^{-1}$ (viz.\ 
adiabatic), though small compared to the thermal relaxation time
$\tau$ (viz.\ rapid). In the usual case, $\mathbf{F}(t)$ remains
within a plane that includes the origin:
$\mathbf{F}(t) = b\,\hat{\mathbf{x}}+ at\,\hat{\mathbf{z}}$,
with $-T_{0}/2 < t < T_{0}/2$, and $\omega_{0}^{-1}\ll T_{0} \ll \tau$. 

ARP can be used to implement a NOT gate on the quantum state of a qubit.
If one identifies the computational basis states $|0\rangle$ and $|1\rangle$, 
respectively, with the spin-up and spin-down eigenstates along the
initial direction of the external field $\mathbf{F}(-T_{0}/2)$, then ARP maps 
$|0\rangle \leftrightarrow |1\rangle$ which is the defining operation of a 
NOT gate. Occurrence of a transition during ARP corresponds to an error in the 
NOT gate since the Bloch vector is not inverted, and thus
$|i\rangle \rightarrow |i\rangle$, ($i=0,1$). Thus we can identify the ARP 
transition probability with the NOT gate error probability. For ARP, 
the adiabatic nature of the inversion ensures that the error probability is 
exponentially small. The price paid for this reliability, however, is an 
extremely slow NOT gate.

In twisted adiabatic rapid passage, the external field
is allowed to twist around its initial direction with azimuthal angle
$\phi (t)$ at the same time that it is adiabatically inverted: 
$\mathbf{F}(t)
= b\cos\phi (t)\,\hat{\mathbf{x}} + b\sin\phi (t) \,\hat{\mathbf{y}} + at\,
\hat{\mathbf{z}}$. Reference~\onlinecite{bry} showed that for twisted ARP, the
exponentially small transition probability contains a factor $\exp
[\,\Gamma_{g}\, ]$ of purely geometric origin. The simplest case
where $\Gamma_{g}\neq 0$ corresponds to quadratic twist: $\phi (t) = 
Bt^{2}$. Zwanziger et.\ al.\ \cite{zwa} were able to experimentally 
realize ARP with quadratic twist and obtained results in agreement 
with the predictions of Reference~\onlinecite{bry}.

In this paper we will consider twisted rapid passage with polynomial twist,
$\phi (t) \sim Bt^{n}$, and we will focus exclusively on qubit inversions
done at \textit{non-adiabatic}\/ rates.
Although we will briefly consider quadratic twist in Section~\ref{sec2} as a 
test case for our numerical simulations, our interest will \textit{not}\/ be 
the geometric effect of Reference~\onlinecite{bry}. Instead, our primary focus 
will be on establishing the existence of 
multiple avoided crossings during twisted rapid passage when $n\geq 3$,
and with exploring some of their consequences. After
general considerations (Section~\ref{sec2}), we will explicitly 
examine cubic ($n=3$) and 
quartic ($n=4$) twist, and will provide clear evidence that the 
multiple avoided crossings produce strong interference  effects in the 
qubit transition probability. The transition probability is shown to be 
a function of the twist strength $B$, which can be
used to control the time-separation of the avoided crossings, and hence 
the character of the interference (constructive or destructive). 
Cubic and quartic twist are examined in Sections~\ref{sec3} and 
\ref{sec4}, respectively. We shall see that interference between the 
multiple avoided crossings can greatly enhance or suppress qubit 
transitions. The interference effects are a direct consequence of the 
temporal phase coherence of the wavefunction. The ability to vary this 
coherence by varying the temporal separation of the avoided crossings renders 
twisted rapid passage with adjustable twist strength into a temporal 
interferometer through which qubit transitions can be controlled. It will be 
shown that quartic twist can implement qubit inversion
non-adiabatically while operating at a fidelity that exceeds the threshold for
fault tolerant operation. Finally, in Section~\ref{sec5}, we 
summarize our results and discuss possible application of this interference 
mechanism to quantum computing. In particular, we describe how one might
use non-adiabatic rapid passage with quartic twist to construct a fast 
fault-tolerant quantum CNOT gate.

It is worth noting that experimental \textit{confirmation\/} of the work 
described in this paper has recently been carried out by Zwanziger et.\ al.\ 
\cite{jwz}. They have realized \textit{non}-\textit{adiabatic} rapid 
passage with both cubic and quartic twist, and have observed clear evidence of 
constructive and destructive interference in the qubit transition probability 
due to interference between the avoided crossings, with excellent agreement 
between the experimental data and our numerical simulations. This experimental 
work provides clear proof-of-principle for our thesis that controllable
quantum interference exists during twisted rapid passage. With this 
thesis now experimentally confirmed, future research can focus on applying 
this interference to the task of constructing fast fault tolerant quantum CNOT 
and NOT gates.

After this paper was submitted, previous work was brought to our
attention which also examined models of rapid passage in which more than
one avoided crossing is possible, and in which interference effects 
were also considered \cite{lim,suo,joy}. These papers focused solely on the 
adiabatic limit. Application of this adiabatic theory to the Zwanziger 
experiment yields predictions that are in poor agreement with the experimental
results \cite{jwz}. This failing is no doubt a consequence of the non-adiabatic 
character of this experiment whose results thus lies beyond the scope
of the adiabatic theory developed in these papers. In contrast, the work we 
present below is principally interested in the non-adiabatic limit, and our 
simulation results are in full agreement with experiment \cite{jwz}. We
also consider possible application of these interference effects to the 
construction of \textit{fast} \textit{fault}-\textit{tolerant} quantum CNOT 
and NOT gates. Ref.~\cite{lim,suo,joy} do not consider such applications.

\section{\label{sec2}Twisted Rapid Passage}

We begin by briefly summarizing the essential features of rapid
passage in the absence of twist. Twistless rapid passage describes 
a wide variety of phenomena, ranging from magnetization reversal in
NMR, to electronic transition during a slow atomic collision. The
essential situation is that of a qubit which is Zeeman-coupled to a 
background field $\mathbf{F}(t)$,
\begin{equation}
H(t) = \mbox{\boldmath $\sigma$}\cdot\mathbf{F}(t) =
         \left( \begin{array}{cc}
                   at & b \\
                   b & -at
                \end{array} \right) \hspace{0.1in} ,
\label{labham}
\end{equation}
with $\mathbf{F}(t) = b\,\hat{\mathbf{x}} + at\,
\hat{\mathbf{z}}$. This particular form for $\mathbf{F}(t)$ 
describes inversion of
the background field in such a way that it remains in the x-z plane
throughout the inversion. For simplicity, we assume $a,\: b > 0$
throughout this paper. The instantaneous energies 
$E_{\pm}(t)$ are:
\begin{equation}
E_{\pm}(t) = \pm\sqrt{b^{2} + (at)^{2}} \hspace{0.1in} ,
\end{equation}
and an avoided crossing is seen to occur at $t=0$ where the energy
gap is minimum. The Schrodinger dynamics 
for twistless rapid passage can be solved exactly for arbitrary
values of $a$ and $b$ \cite{lan,zen}, and yields the Landau-Zener 
expression for the transition probability $P_{LZ}$:
\begin{equation}
P_{LZ} = \exp\left[\, -\frac{\pi b^{2}}{\hbar |a|}\,\right] 
           \hspace{0.1in} .
\label{plz}
\end{equation}

\subsection{Twisted Rapid Passage and Multiple Avoided Crossings}

In {\it twisted\/} rapid passage, the background field 
$\mathbf{F}(t)$ 
is allowed to twist around its initial direction during the course
of its inversion: $\mathbf{F}(t) = b\cos\phi (t)\,\hat{\mathbf{x}}
+b\sin\phi (t)\,\hat{\mathbf{y}} + at\, \hat{\mathbf{z}}$. It proves 
convenient to transform to the rotating frame in which the x-y
component of the background field is instantaneously at rest. This is 
accomplished via the unitary transformation $U(t) = \exp [-(i/2)\phi
(t)\sigma_{z}]$. The Hamiltonian $\overline{H}(t)$ in this frame is:
\begin{equation}
\overline{H}(t) 
      = \mbox{\boldmath $\sigma$}\cdot\overline{\mathbf{F}}
= \left( \begin{array}{cc}
                 \left(\, at -\frac{\hbar}{2}\dot{\phi}\,\right)
                          & b \\
                 b & 
                     -\left(\, at - \frac{\hbar}{2}\dot{\phi}\,\right)
              \end{array} \right) 
\hspace{0.1in} ,
\label{rotham}
\end{equation}
where $\overline{\mathbf{F}}(t) = b\,\hat{\mathbf{x}}+ (at-\hbar\dot{\phi}/2)
\,\hat{\mathbf{z}}$ is the background field as seen in the rotating frame,
and a dot over a symbol represents the time derivative of that symbol.
The instantaneous energy eigenvalues are $\overline{E}_{\pm}(t) = \pm 
\sqrt{\left(\, at - (\,\hbar\dot{\phi}\,)/2\,\right)^{2} + 
b^{2}}$. Avoided crossings occur when the energy gap is minimum, 
corresponding to when 
\begin{equation}
at - \frac{\hbar}{2}\,\frac{d\phi}{dt} = 0 \hspace{0.1in} .
\label{cond}
\end{equation}
For polynomial twist: $\phi_{n}(t) = c_{n}Bt^{n}$, where $B$ is the
twist strength. The dimensionless constant $c_{n}$ has been introduced
to simplify some of the formulas below. For later convenience, we 
chose $c_{n} = 2/n$. For polynomial twist, it is easily checked that
eqn.~(\ref{cond}) always has the root:
\begin{equation}
  t = 0 \hspace{0.1in} ,
\label{avdx1}
\end{equation}
and that for $n\ge 3$, eqn.~(\ref{cond}) also has the $n-2$ roots:
\begin{equation}
  t = \left(\, {\rm sgn}\, B\,\right)^{\frac{1}{n-2}}\,
       \left(\,\frac{a}{\hbar |B|}\,\right)^{\frac{1}{n-2}}
         \hspace{0.1in} .
\label{avdx2}
\end{equation}
All together, equation~(\ref{cond}) has $n-1$ roots, 
though only the real roots correspond to avoided crossings. 
For quadratic twist ($n=2$), only eqn.~(\ref{avdx1}) arises. Thus, for
this case, only the avoided crossing at $t=0$ is possible. For $n\geq 3$, 
along with the avoided crossing at $t=0$, real solutions to 
eqn.~(\ref{avdx2}) also occur. The different possibilities for this situation
are summarized in Table~\ref{table1}.
\begin{table*}[h]
\caption{\label{table1}Classification of regimes under which multiple avoided
crossings occur for polynomial twist with $n \geq 3$.}
\begin{ruledtabular}
  \begin{tabular}{lrl}
\multicolumn{3}{c}{1.  $\underline{ sgn\, B = +1}$} \\
   (a)\hspace{0.2in}  $n$ odd;     &  $\:\:\: 2$ avoided crossings at:  &  
$\:\:\: t=0$ and $t=\left(a/\hbar B\right)^{1/(n-2)}$ \\
   (b)\hspace{0.2in}  $n$ even;  &  $\:\:\: 3$ avoided crossings at: &  
$\:\:\: t=0$  and $t=\pm\left(a/\hbar B\right)^{1/(n-2)}$ \\
\multicolumn{3}{c}{2.  $\underline{ sgn\, B = -1}$}  \\
   (a)\hspace{0.2in}  $n$ odd;     &   $\:\:\: 2$ avoided crossings at: &
$\:\:\: t=0$  and $t=-\left(a/\hbar |B|\right)^{1/(n-2)}$ \\
   (b)\hspace{0.2in}  $n$ even;  & $\:\:\: 1$ avoided crossing at: & 
$\:\:\: t=0$ \\
   \end{tabular}
\end{ruledtabular}
\end{table*} 
We see that for polynomial twist with $n\geq 3$, multiple avoided crossings
always occur for positive twist strength $B$, while for negative
twist strength, multiple avoided crossings only occur when $n$ is odd.
Note that the time separating the multiple avoided crossings can be adjusted
by variation of the twist strength $B$ and/or the inversion rate $a$.

\subsection{Brief Detour: Background on Quadratic Twist}

As mentioned earlier, quadratic twist has already been examined in the
literature \cite{bry}. It is of interest here only because its dynamics can
be solved exactly, and thus allows us to test our numerical simulations
before proceeding to unexplored cases of twisted rapid
passage. For quadratic twist, $\dot{\phi}_{2} = 2Bt$. Inserting this into
eqn.~(\ref{rotham}) gives $\overline{\mathbf{F}}(t) = b\,\hat{\mathbf{x}}+
\overline{a} t\,\hat{\mathbf{z}}$, with $\overline{a}= a -\hbar |B|\, 
(\mathrm{sgn}\, B)$.
Thus rapid passage with quadratic twist maps onto twistless rapid passage
with $a\rightarrow \overline{a}$. This allows us to obtain an exact result 
for the transition probability $P_{2}$ for arbitrary values of $a$ and $b$
from eqn.~(\ref{plz}) with $a\rightarrow \overline{a}$:
\begin{equation}
P_{2} = \exp\left[\, -\frac{\pi b^{2}}{ \hbar |\, a - \hbar |B|\, (
         \mathrm{sgn}\, B)\, | } \right] \hspace{0.1in} .
\label{tranprob}
\end{equation}
In the adiabatic limit, this reduces to $P_{2}=P_{LZ}\exp [\Gamma_{g}]$,
where $\Gamma_{g}=-\pi Bb^{2}/a^{2}$ is the geometric exponent discovered
in Ref.~\cite{bry}. Eqn.~(\ref{tranprob}) makes the interesting prediction 
that a \textit{complete\/} quenching of transitions will occur when
$\mathrm{sgn}\, B = +1$ and $a = \hbar B$, while no such quenching is possible
for $\mathrm{sgn}\, B = -1$. Zwanziger et.~al.\ \cite{zwa} were able to 
realize rapid passage with quadratic twist experimentally and confirmed the
existence of $\Gamma_{g}$, and the twist-dependent quenching of transitions.
We now show that our  numerical simulation also reproduces these effects.

\subsection{Simulation Details}

The equations that drive the numerical simulation follow from the
Schrodinger equation in the non-rotating frame:
\begin{equation}
i\hbar\frac{\partial}{\partial t}\, |\psi\rangle = H(t)\, |\psi\rangle
                 \hspace{0.1in} ,
  \label{schro}
\end{equation}
where $H(t) = \mbox{\boldmath $\sigma$}\cdot\mathbf{F}(t)$, and 
$\mathbf{F}(t) = b\cos\phi (t)\,\hat{\mathbf{x}} + b\sin\phi (t)\,
\hat{\mathbf{y}} + at\,\hat{\mathbf{z}}$. To obtain these equations in
the adiabatic representation, we expand $|\psi (t)\rangle$ in the 
instantaneous eigenstates $|E_{\pm}(t)\rangle$ of $H(t)$:
\begin{equation}
|\psi (t)\rangle = S(t)\; e^{-\frac{i}{\hbar}\int_{-T_{0}/2}^{t}\, d\theta\,
              \left( E_{-}-\hbar\dot{\gamma}_{-}\right)}\;
               |E_{-}(t)\rangle \; - \;
           I(t)\; e^{-\frac{i}{\hbar}\int_{-T_{0}/2}^{t}\, d\theta\,
              \left( E_{+} -\hbar\dot{\gamma}_{+}\right)}\;
               |E_{+}(t)\rangle \hspace{0.1in} .
\label{expan}
\end{equation}
Here $\gamma_{\pm}(t)$ are the geometric phases \cite{geo} associated with
the energy-levels $E_{\pm}(t)$, respectively, and
\begin{equation}
\dot{\gamma}_{\pm}(t) = i\langle\, E_{\pm}(t)\, |\,\frac{d}{dt}\, |\, 
                         E_{\pm}(t)\, \rangle = 
     i \langle\, E_{\pm}(t)\, |\,\dot{E}_{\pm}(t)\,\rangle
                         \hspace{0.1in} .
\label{berfaz}
\end{equation}
Substituting eqn.~(\ref{expan}) into (\ref{schro}), and using the
orthonormality of the instantaneous eigenstates, one obtains the
equations of motion for the expansion coefficients $S(t)$ and $I(t)$:
\begin{subequations}
\label{alleom}
\begin{eqnarray}
\frac{dS}{dt} & = & -\Gamma^{\ast}(t)\; e^{-i\int_{-T_{0}/2}^{t}\,d\theta\,
                        \delta
                  (\theta )}\; \vspace{0.15in}I(t) \hspace{0.1in} , 
                     \label{eoma} \\
\frac{dI}{dt} & = & \mbox{} \;\;\; \Gamma (t)\; e^{i\int_{-T_{0}/2}^{t}\, 
                      d\theta\,
                     \delta (\theta )} \; S(t) \hspace{0.1in} . \label{eomb}
\end{eqnarray}
\end{subequations} 
Here,
\begin{eqnarray}
\delta (t) & = & \frac{E_{+}(t) - E_{-}(t)}{\hbar} - \left(\,
                      \dot{\gamma}_{+}(t)
                   - \dot{\gamma}_{-}(t) \,\right) \hspace{0.1in} ,
                      \label{delta} \\
\Gamma (t) & = & \langle\, E_{+}(t)\, |\,\dot{E}_{-}(t) \,\rangle 
                    \hspace{0.1in} , 
                   \label{Gamma}
\end{eqnarray}
and one can show that $\Gamma^{\ast}(t) = -\,\langle E_{-}(t)\, |\,
\dot{E}_{+}(t) \,\rangle$.
Eqns.~(\ref{alleom}) are the qubit equations of motion in the adiabatic
representation and include the influence of the geometric phase on the dynamics
through $\delta (t)$. In the case of twistless rapid passage, the geometric
phase vanishes, and eqns.~(\ref{alleom}) reduce to the well-known equations
of motion for a two-level system found in Ref.~\cite{thr}. 
Eqns.~(\ref{alleom}) can be put in 
dimensionless form if we introduce the dimensionless variables: 
$\tau = (a/b)t$, $\overline{\Gamma} = (b/a)\Gamma$, and $\overline{\delta} = 
(b/a)\delta$. Here $a$ and $b$ are the parameters that appear in the
background field $\mathbf{F}(t)$. One obtains:
\begin{subequations}
\label{alldeom}
\begin{eqnarray}
\frac{dS}{d\tau } & = & -\overline{\Gamma}^{\ast}\; e^{-i
                          \int_{-\tau_{0}/2}^{\tau }\, d\theta\,
                           \overline{\delta}(\theta )}\; I(\tau )
                            \vspace{0.25in}\hspace{0.1in} , \label{deoma} \\
\frac{dI}{d\tau } & = & \mbox{}\;\;\;\overline{\Gamma}\, e^{i
                         \int_{-\tau_{0}/2}^{\tau }\,
                         d\theta \,\overline{\delta}(\theta )}\; S(\tau )
                          \hspace{0.1in} , \label{deomb}
\end{eqnarray}
\end{subequations}
where $\tau_{0} = (a/b)T_{0}$ is the (dimensionless) time over which
the qubit evolves. For rapid passage, the qubit is initially in the
negative energy level $|E_{-}(-\tau_{0}/2)\,\rangle$. This corresponds to
the initial condition:
\begin{subequations}
\label{allinicond}
\begin{eqnarray}
S(-\tau_{0}/2) & = & 1 \hspace{0.1in} , \label{iniconda} \\
I(-\tau_{0}/2) & = & 0 \hspace{0.1in} . \label{inicondb}
\end{eqnarray}
\end{subequations}
Our numerical simulation integrates eqns.~(\ref{alldeom}) over the
time-interval $[\,-\tau_{0}/2,\;\tau_{0}/2\,]$ subject to initial
condition (\ref{allinicond}). From this we determine the asymptotic
transition probability $P$:
\begin{equation}
P = |I(\tau_{0}/2)|^{2} \hspace{0.1in} ,
\label{tranprob2}
\end{equation}
for $\tau_{0} \gg 1$. Later, we will need the $\tau$-values corresponding to
the avoided crossings. These are determined by rewriting eqns.~(\ref{avdx1}) 
and (\ref{avdx2}) in dimensionless form. To this end, we introduce
\begin{equation}
\eta_{\,n} = \frac{\hbar B \, b^{n-2}}{a^{n-1}} \hspace{0.1in} ,
\label{twststr}
\end{equation}
and recalling that $\tau = (a/b)t$, one easily obtains:
\begin{equation}
\tau = 0 \hspace{0.1in} ,
\label{davdx1}
\end{equation}
and
\begin{equation}
\tau = \left(\, sgn\,\eta_{\,n}\,\right)^{\frac{1}{n-2}}\,
        \left[\,\frac{1}{|\eta_{\,n}|}\,\right]^{\frac{1}{n-2}}
         \hspace{0.1in} .
\label{davdx2}
\end{equation}
The avoided crossings correspond to $\tau = 0$ and also, for $n\geq 3$,
the real solutions of eqn.~(\ref{davdx2}).

\subsection{Simulation Test Case: Quadratic Twist}

\begin{figure}[h]
\includegraphics[scale=0.5]{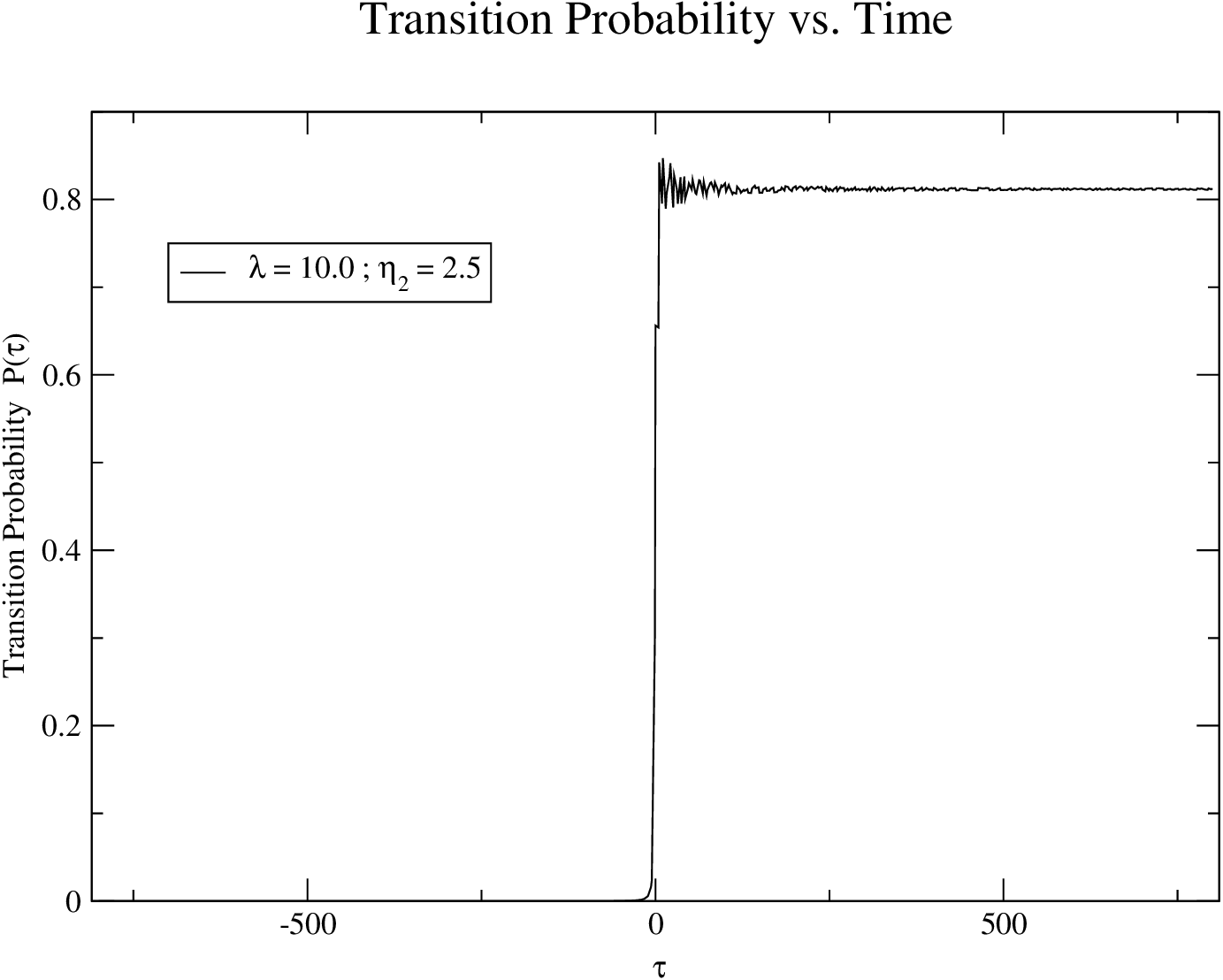}
\caption{\label{fig1}Representative plot of transition probability
$P(\tau )$ for quadratic twist with $\lambda = 10.0$ and $\eta_{\, 2} = 2.5$.}
\end{figure}

\begin{figure}[h]
\includegraphics[scale=0.5]{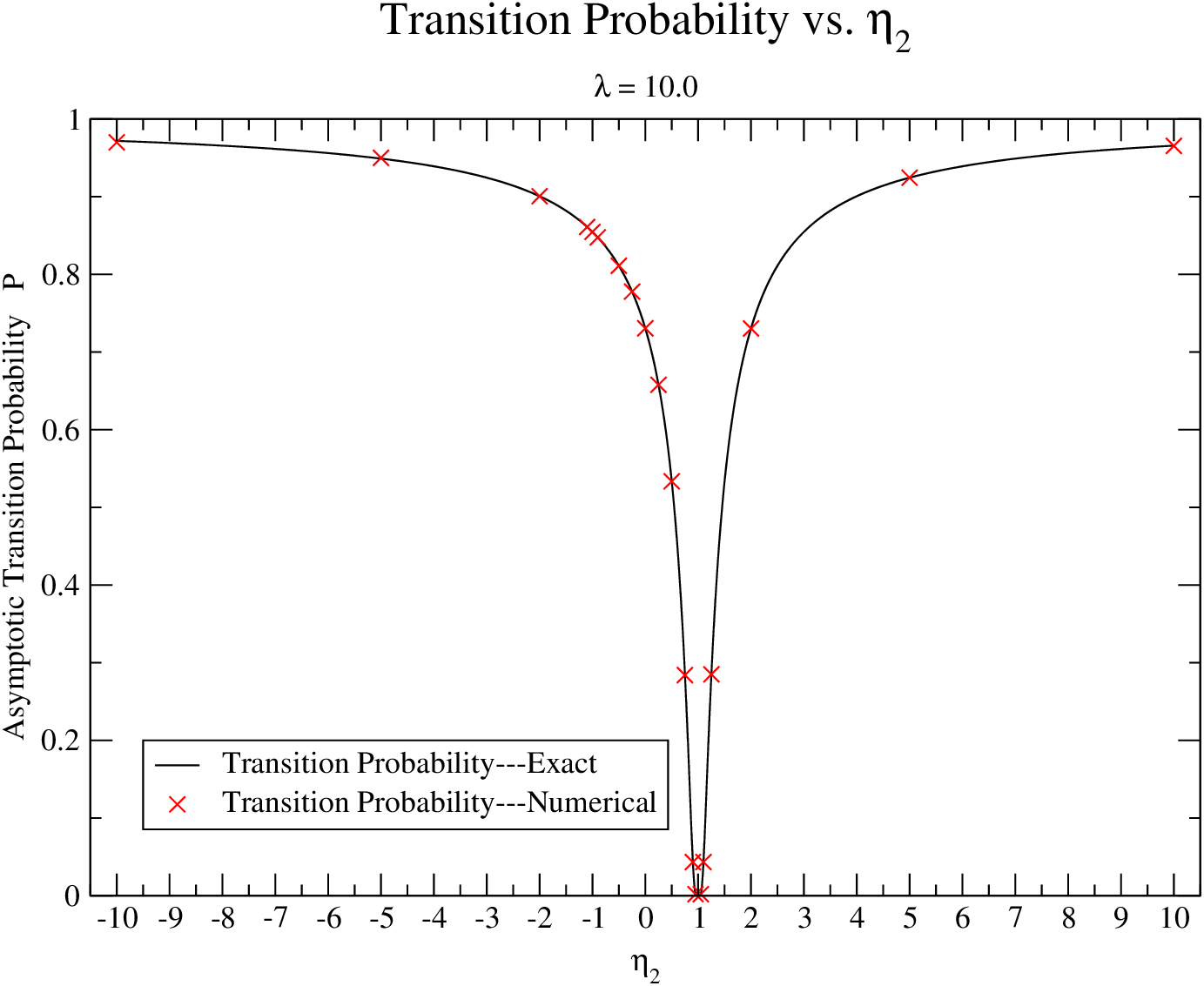}
\caption{\label{fig2}Numerical results for the asymptotic transition 
probability $P$ versus $\eta_{\, 2}$ for quadratic twist with 
$\lambda = 10.0$. Also plotted is the exact result $P_{2}$.}
\end{figure}

\begin{figure}[h]
\includegraphics[scale=0.5]{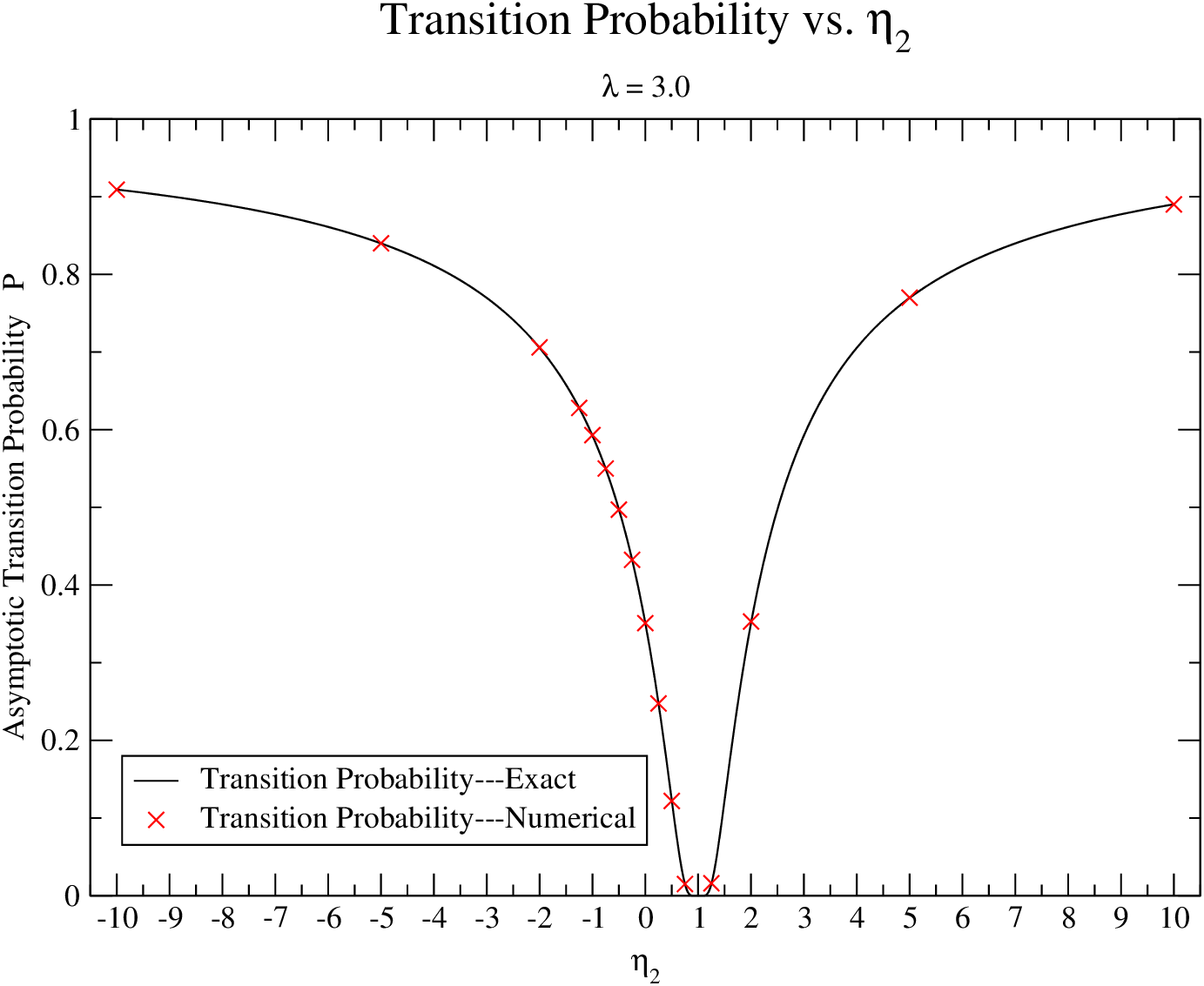}
\caption{\label{fig3}Numerical results for the asymptotic transition 
probability $P$ versus $\eta_{\, 2}$ for quadratic twist with 
$\lambda = 3.0$. Also plotted is the exact result $P_{2}$.}
\end{figure}

For quadratic twist $\phi_{2}(t) = Bt^{2}$. The instantaneous eigenvalues
and eigenvectors of $H(t)$ are easily found to be $E_{\pm}(t)=\pm E(t)$,
where $E(t)=\sqrt{b^{2}+(at)^{2}}$, and
\begin{equation}
|E_{+}(t)\,\rangle = \left( \begin{array}{c}
                               \cos\frac{\theta}{2} \\
                               \sin\frac{\theta}{2}\, e^{i\phi_{2}}
                            \end{array} \right) \hspace{0.5in} ;
                     \hspace{0.5in} |E_{-}(t)\,\rangle =
                          \left( \begin{array}{c}
                                    \sin\frac{\theta}{2} \\
                                    -\cos\frac{\theta}{2} \, e^{i\phi_{2}}
                                 \end{array} \right) \hspace{0.1in} ,
\label{eigstates}
\end{equation}
with $\cos\theta = at/E$. From the eigenstates one obtains:
\begin{subequations}
 \label{allmisc}
 \begin{eqnarray}
    \dot{\gamma}_{\pm}(t) & = & -\frac{\dot{\phi}_{2}}{2}\left(\, 1 \mp
                                    \cos\theta\, \right) \hspace{0.1in} ;
                                     \label{gamdot} \\
    \Gamma (t) & = & \frac{\dot{\theta}}{2} - i\frac{\dot{\phi}_{2}}{2}
                       \sin\theta \hspace{0.1in} ; \label{biggam} \\
    \delta (t) & = & \frac{2E}{\hbar} - \dot{\phi}_{2}\cos\theta 
                       \hspace{0.1in} . \label{delt} 
 \end{eqnarray}
\end{subequations}
$\overline{\Gamma}(\tau )$ and $\overline{\delta}(\tau )$ are then
determined from eqns.~(\ref{biggam}) and (\ref{delt}) and are found
to depend parametrically on the dimensionless ``inversion rate''
$\lambda = \hbar a/b^{2}$ and the dimensionless ``twist strength''
$\eta_{\, 2} = \hbar B/a$. ``Inversion rate'' and ``twist strength'' are 
placed in quotes as $\lambda$ does not depend solely on the inversion rate 
$a$, nor $\eta_{\, 2}$ solely on the twist strength $B$. Crudely speaking,
$\lambda = 1$ can be thought of as the boundary separating adiabatic and
non-adiabatic inversion rates, with $\lambda > 1$ corresponding to
non-adiabatic inversion. Having determined
$\overline{\Gamma}(\tau )$ and $\overline{\delta}(\tau )$, 
eqns.~(\ref{alldeom}) are integrated numerically using an adjustable
step-size fourth-order Runge-Kutta algorithm. To simplify comparison
of the numerical result for the transition probability with the exact
result $P_{2}$, we re-write eqn.~(\ref{tranprob}) in terms of $\lambda$ 
and $\eta_{\, 2}$. One finds:
\begin{equation}
P_{2} = \exp\left[\, -\frac{\pi}{\lambda}\,
                       \frac{1}{\left|\, 1 - \eta_{\, 2}\,\right| } \,\right]
                         \hspace{0.1in} .
\label{dtranprob}
\end{equation}
Figure~\ref{fig1} shows a representative plot of the transition probability
$P(\tau ) = |I(\tau )|^{2}$ versus $\tau$. 
It is clear for the Figure that 
the transition occurs in the vicinity of the avoided crossing at $\tau = 0$.
Note also that $P(\tau )$ has a small oscillation about its asymptotic
value $P = \lim_{\tau\rightarrow\infty}\, P(\tau )$. To average out the 
oscillation, $P(\tau )$ (for given $\lambda$ and $\eta_{\, 2}$) was 
calculated for 10 different values of $\tau\gg 1$, and $P$ was identified with 
the average. 
Figures~\ref{fig2} and \ref{fig3} show our numerical results for $P$ for
various values of $\eta_{\, 2}$ for $\lambda = 10.0$ and $\lambda = 3.0$, 
respectively. Also plotted in each of these Figures is the exact result 
$P_{2}$ (eqn.~(\ref{dtranprob})). 
Figures~\ref{fig2} and \ref{fig3} show that our numerical results are in 
excellent agreement with the exact result $P_{2}$, and clearly show the 
quenching of transitions at $\eta_{\, 2} = 1$, and the absence of quenching 
for negative $\eta_{\, 2}$. The $\lambda$ values shown are purposely highly 
non-adiabatic. We
see that the twist-induced quenching clearly persists into the non-adiabatic
regime, although the width of the quench decreases with increasing $\lambda$. 
The agreement of our simulations with eqn.~(\ref{dtranprob}) at small 
$\eta_{\, 2}$ indicates that our simulations also account for the geometric 
factor $\exp\left[\Gamma_{g}\right]$ in $P_{2}$. Having established that
our numerical algorithm correctly reproduces the essential results of
rapid passage with quadratic twist, we go on to consider the unexplored 
areas of rapid passage with higher order twist. Referring to 
Table~\ref{table1}, we see that all cases with odd $n$ have 2 avoided
crossings. Cubic ($n=3$) twist corresponds to the simplest example
of odd-order twist, and it is examined in the following Section.
Similarly, quartic ($n=4$) twist is the simplest example of even-order twist,
and we examine it in Section~\ref{sec4}. 

\section{\label{sec3}Cubic Twist}

\begin{figure}[h]
\includegraphics[scale=0.5]{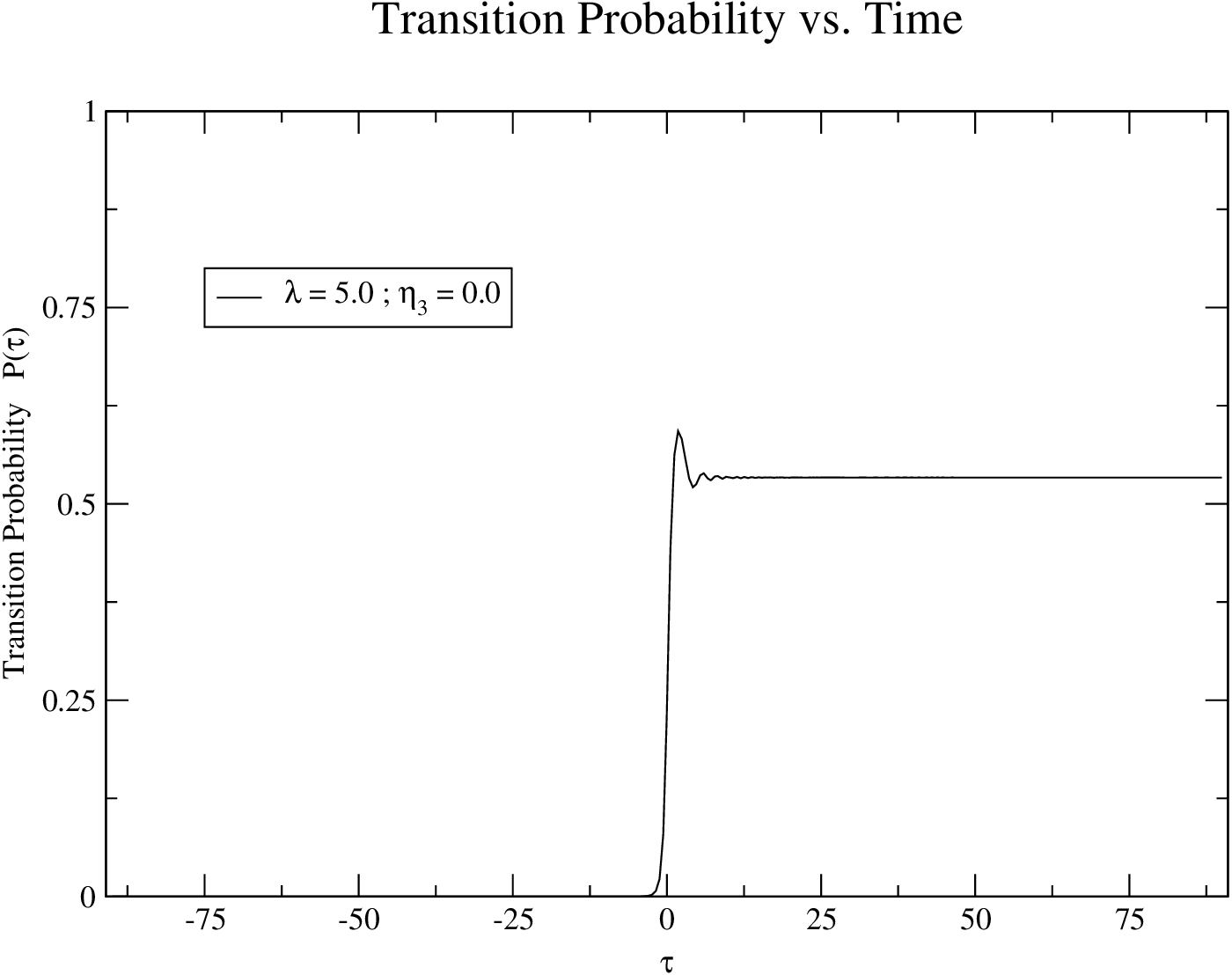}
\caption{\label{fig4} Plot of the transition probability $P(\tau )$ for
twistless non-adiabatic rapid passage with $\lambda = 5.0$ and $\eta_{\, 3} 
= 0$.}
\end{figure}

Having successfully tested our numerical algorithm against the exact
results for quadratic twist, we go on to consider cubic twist for which
$\phi_{3}(t)=(2/3)Bt^{3}$, and $\eta_{\, 3} = \hbar B b/a^{2}$ (see
eqn.~(\ref{twststr})). As in Section~\ref{sec2}, the instantaneous
eigenvalues of $H(t)$ are $E_{\pm}(t) =\pm E(t)$, and the instantaneous
eigenstates are given by eqn.~(\ref{eigstates}) with $\phi_{2}(t)\rightarrow
\phi_{3}(t)$. 
Eqns.~(\ref{allmisc}) again apply, however $\dot{\phi}_{2}(t)
\rightarrow \dot{\phi}_{3}(t)$, and $\overline{\Gamma}(\tau )$ and
$\overline{\delta}(\tau )$ are determined from eqns.~(\ref{biggam}) and 
(\ref{delt}). Having determined $\overline{\Gamma}(\tau )$ and
$\overline{\delta}(\tau )$, eqns.~(\ref{alldeom}) can be numerically 
integrated subject to the initial condition specified in 
eqns.~(\ref{allinicond}).
Before examining results of that integration, we show in Figure~\ref{fig4}
a plot of the numerical results for the transition probability $P(\tau )$
for $\lambda = 5.0$ and $\eta_{\, 3} = 0$.
This corresponds to twistless non-adiabatic rapid passage, and we include
this plot for later comparison with related plots for cubic and quartic twist. 
The asymptotic transition probability for this case is $P=0.533$. Thus, if we
were to use this example of twistless non-adiabatic rapid passage to implement
a fast NOT-operation on a qubit, the operation would be slightly more likely 
to produce an inversion (bit-flip) error than not. We will show below that if 
a small amount 
of cubic twist is included, the bit-flip error probability can be reduced by
2 orders of magnitude while still maintaining the non-adiabatic inversion
rate $\lambda = 5.0$. This substantial reduction in error probability is due
to destructive interference between the two avoided crossings that occur 
during rapid passage with cubic twist.

\subsection{Demonstration of Quantum Interference}

\begin{figure}[h]
\includegraphics[scale=0.5]{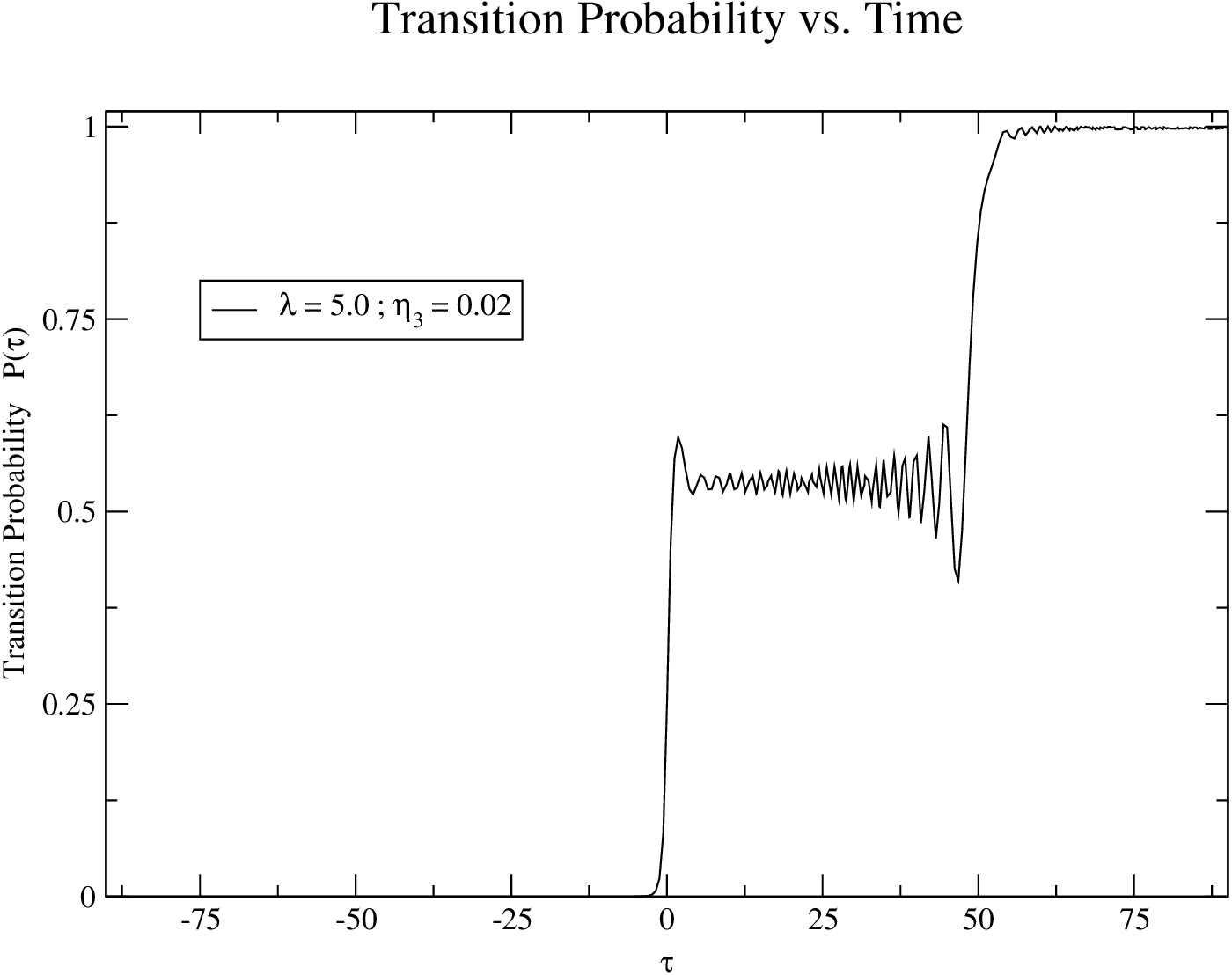}
\caption{\label{fig5} The transition probability $P(\tau )$ for non-adiabatic
rapid passage with cubic twist with $\lambda = 5.0$ and $\eta_{\, 3} = 
0.02$.}
\end{figure}

\begin{figure}[h]
\includegraphics[scale=0.5]{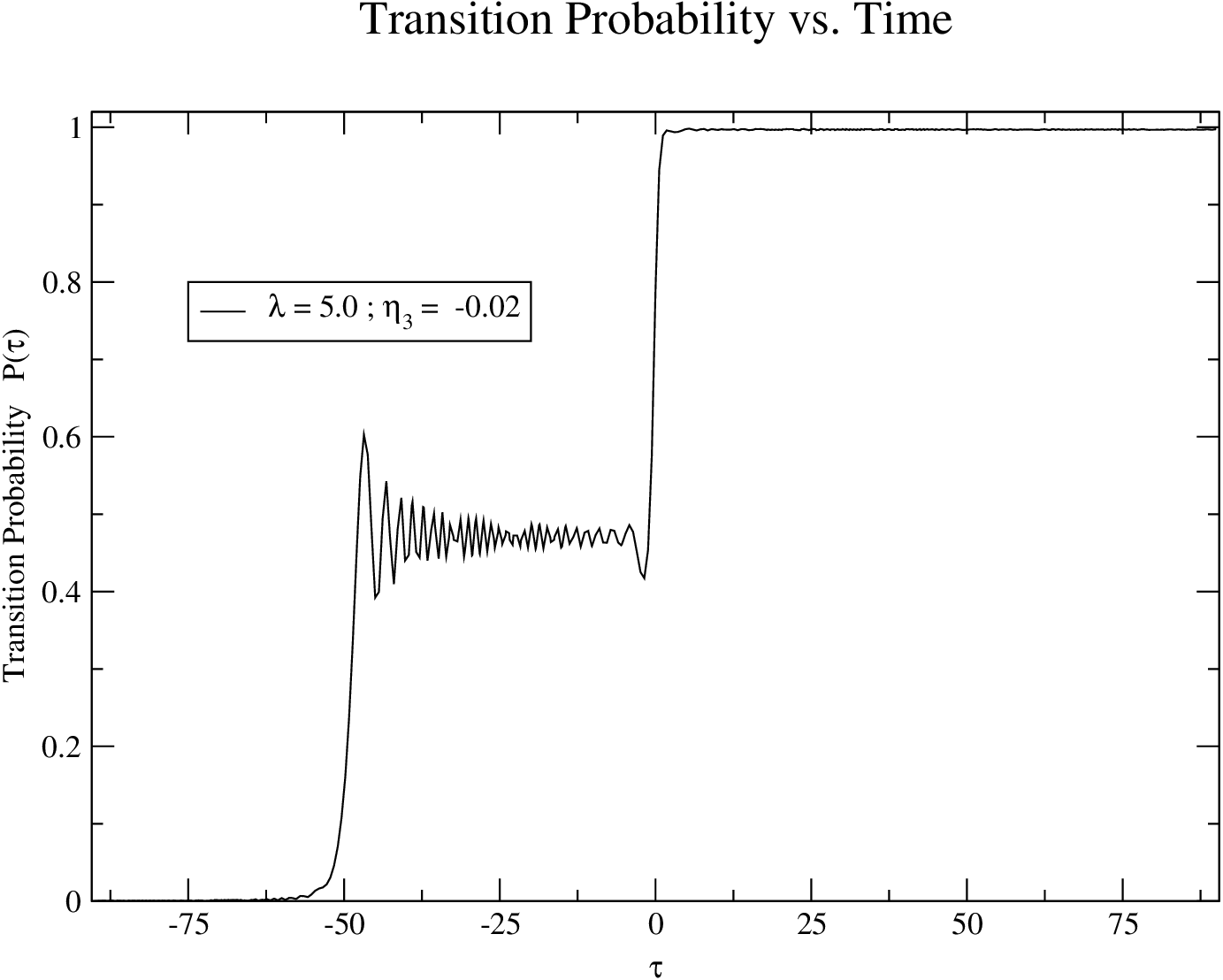}
\caption{\label{fig6} The transition probability $P(\tau )$ for non-adiabatic
rapid passage with cubic twist with $\lambda = 5.0$ and $\eta_{\, 3} = 
-0.02$.}
\end{figure}

\begin{figure}[h]
\includegraphics[scale=0.5]{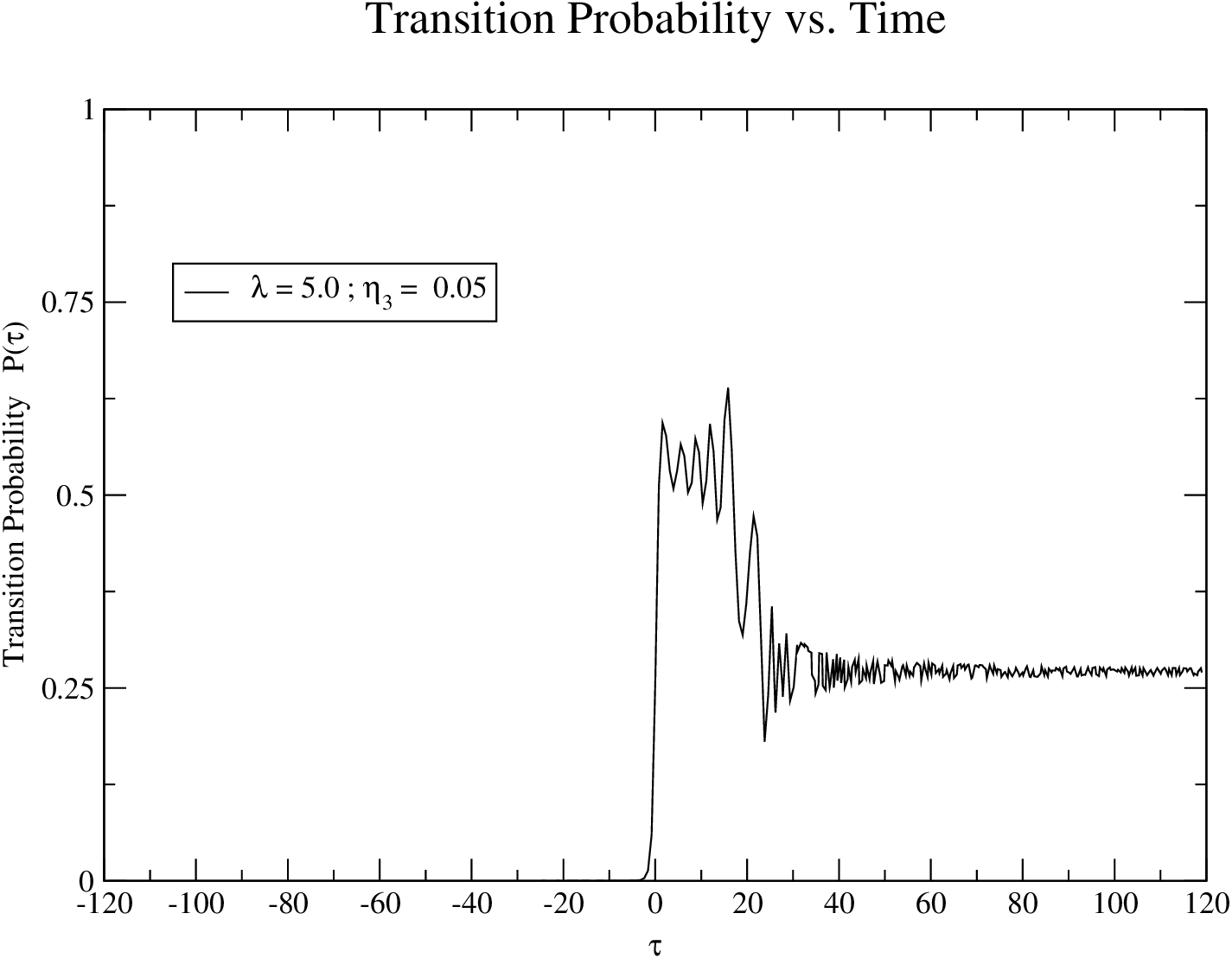}
\caption{\label{fig7} The transition probability $P(\tau )$ for non-adiabatic
rapid passage with cubic twist with $\lambda = 5.0$ and $\eta_{\, 3} = 
0.05$.}
\end{figure}

\begin{figure}[h]
\includegraphics[scale=0.5]{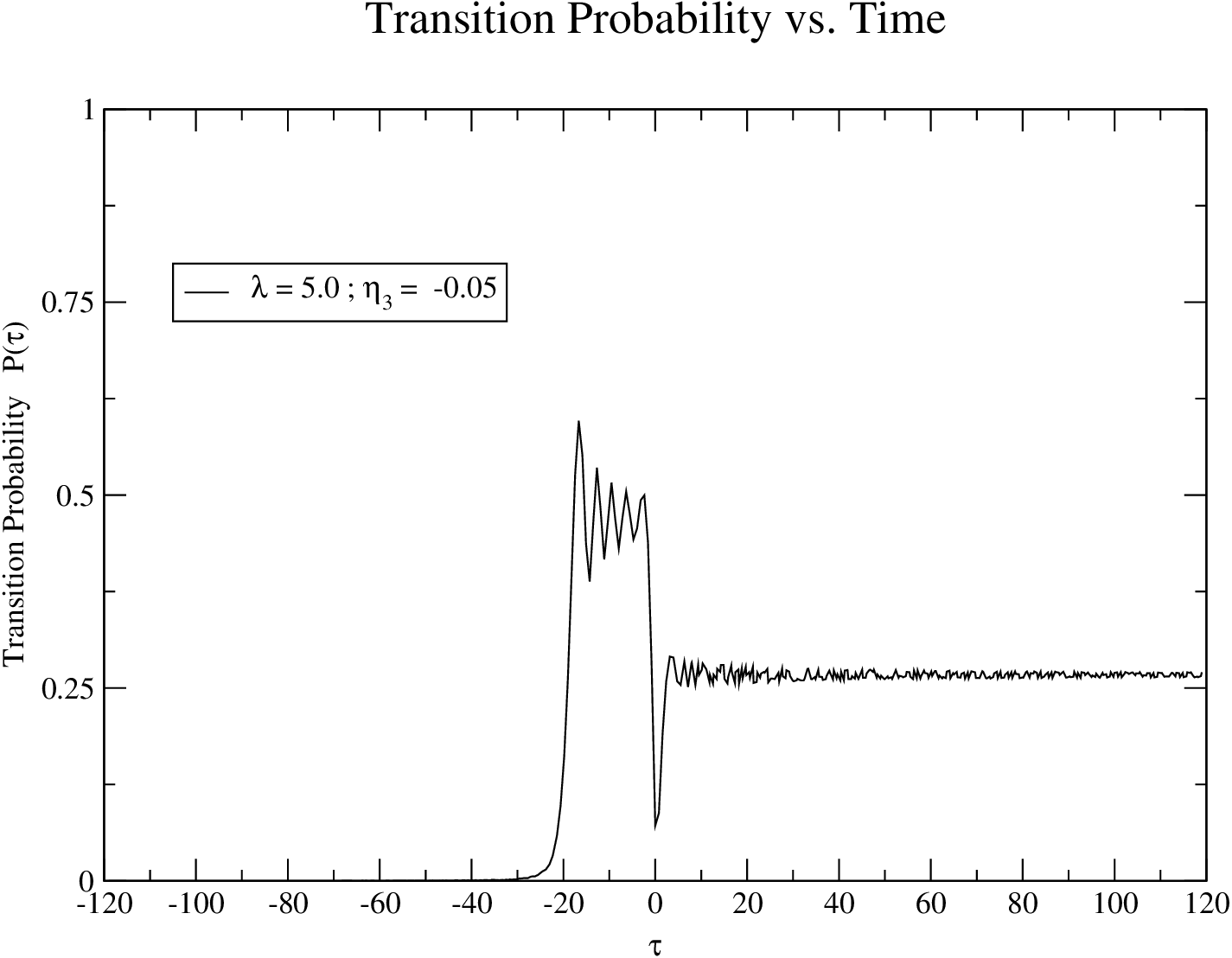}
\caption{\label{fig8} The transition probability $P(\tau )$ for non-adiabatic
rapid passage with cubic twist with $\lambda = 5.0$ and $\eta_{\, 3} = 
-0.05$.}
\end{figure}

From eqns.~(\ref{davdx1}) and (\ref{davdx2}), we see that cubic twist is 
expected to have 2 avoided crossings at $\tau_{1} = 0$ and $\tau_{2}= 
sgn\,\eta_{\, 3} /|\eta_{\, 3} |$.
Figures~\ref{fig5} and \ref{fig6} show $P(\tau )$ for $\lambda = 5.0$
and $\eta_{\, 3} = 0.02$ and $\eta_{\, 3} = -0.02$, respectively. 
Figure~\ref{fig5} (\ref{fig6}) clearly shows the expected avoided crossings
at $\tau = 0$ and $\tau = 50$ ($-50$). It is also clear from these Figures,
and comparison with Figure~\ref{fig4}, that the avoided crossings are
constructively interfering, leading to an asymptotic transition probability
of $P=0.997$. Figures~\ref{fig7} and \ref{fig8} show $P(\tau )$ for
$\lambda = 5.0$ and $\eta_{\, 3} = 0.05$ and $-0.05$, respectively.
The avoided crossings in Figure~\ref{fig7} (\ref{fig8}) clearly occur at 
$\tau = 0$ and $\tau = 20$ ($-20$) as expected. Here the avoided crossings
interfere destructively, with $P = 0.270$. Summarizing, we see 
that: (1) two avoided crossings do occur during rapid passage with cubic 
twist as predicted in Table~\ref{table1}; (2) the avoided crossings 
produce interference effects in the asymptotic transition probability $P$ 
which can be controlled through variation of their separation; and (3) the 
separation of the avoided 
crossings $\Delta\tau_{ac}= |\tau_{2}-\tau_{1}|=1/|\eta_{\, 3}|$ can be 
altered by varying $\eta_{\, 3} = \hbar B b/a^{2}$. We now consider 
two possible applications of this interference effect.

\subsection{Non-Resonant Pump}

\begin{figure}[h]
\includegraphics[scale=0.5]{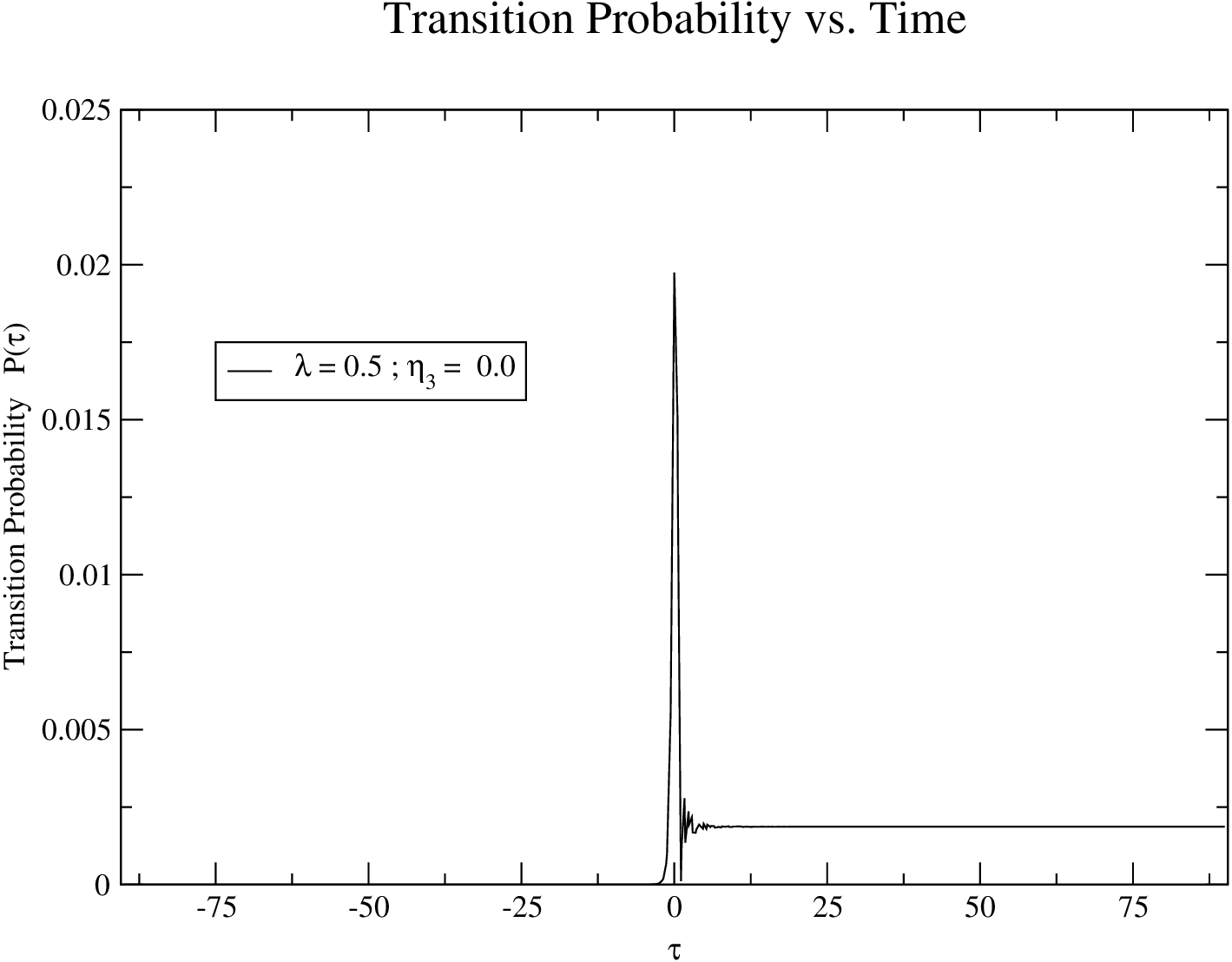}
\caption{\label{fig9} The transition probability $P(\tau )$ for twistless
adiabatic rapid passage with $\lambda = 0.5$ and $\eta_{\, 3} = 0$. Note 
the greatly
reduced vertical scale compared to previous figures.}
\end{figure}

\begin{figure}[h]
\includegraphics[scale=0.5]{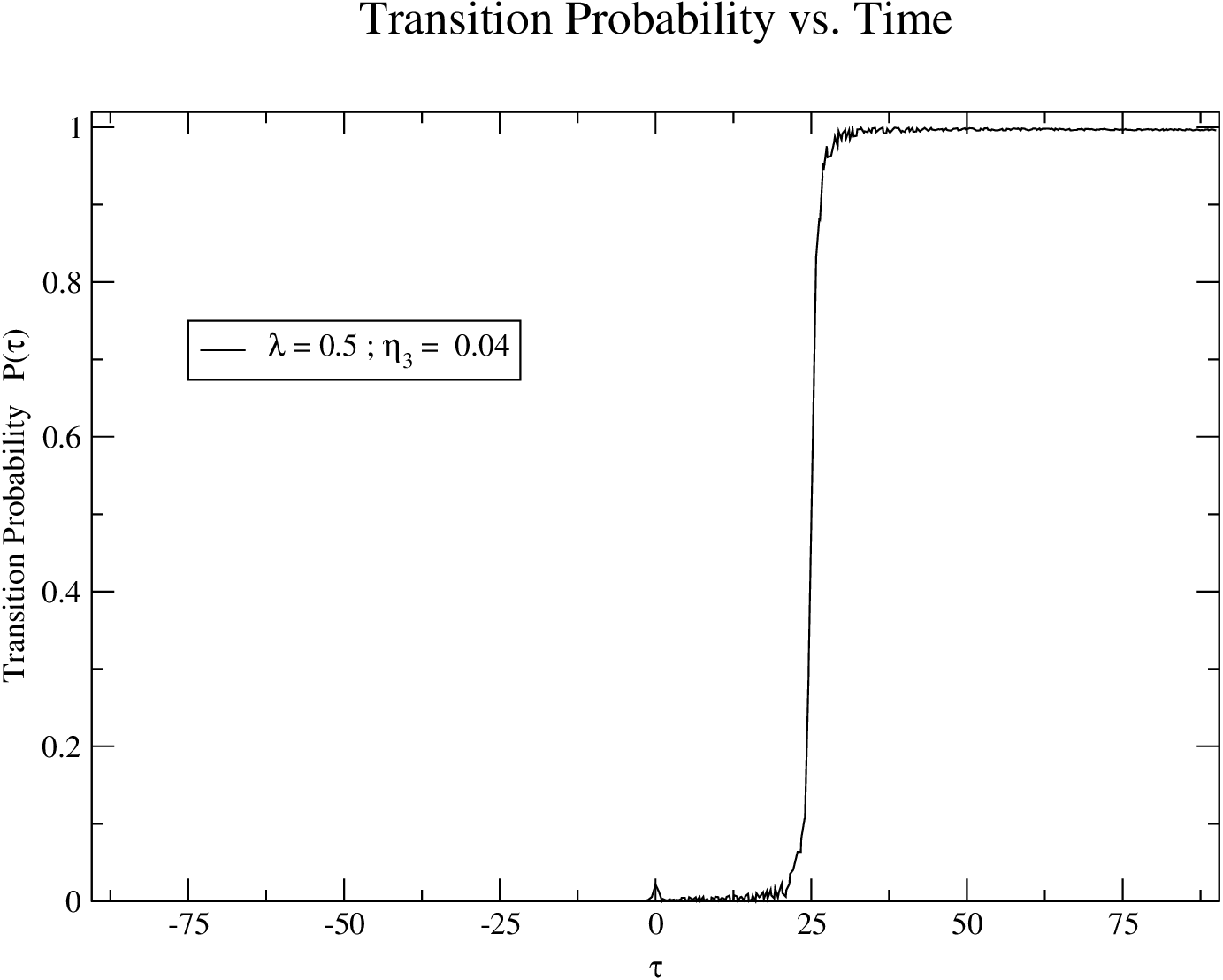}
\caption{\label{fig10} The transition probability $P(\tau )$ for 
adiabatic rapid passage with cubic twist with $\lambda = 0.5$ and 
$\eta_{\, 3} = 0.04$.}
\end{figure}

First, consider twistless adiabatic rapid passage with $\lambda = 0.5$
and $\eta_{\, 3} = 0$. Figure~\ref{fig9} show the transition probability 
$P(\tau )$
for this case. The asymptotic transition probability is 
$P = 1.87\times 10^{-3}$. 
Figure~\ref{fig10} shows $P(\tau )$ for adiabatic
rapid passage with cubic twist with $\lambda = 0.5$ and $\eta_{\, 3} = 0.04$. 
The asymptotic transition probability in this case is $P=0.996$! Thus, by
introducing a small amount of cubic twist, constructive interference between
the avoided crossings transforms adiabatic rapid passage into a non-resonant
pump for the qubit energy levels. Figures~\ref{fig5} and \ref{fig6} indicate 
that, should it be desired, equally large transition probabilities are also 
possible at faster inversion rates $\lambda$. It is worth pointing out that to 
produce such a large transition probability using twistless non-adiabatic 
rapid passage would require $\lambda = 784$ (see eqn.~(\ref{dtranprob}) 
with $\eta_{\, 2} = 0$) as opposed to $\lambda\sim 0.5-5.0$ when cubic twist 
is exploited.

\subsection{Transition Quenching}

\begin{figure}[h]
\includegraphics[scale=0.5]{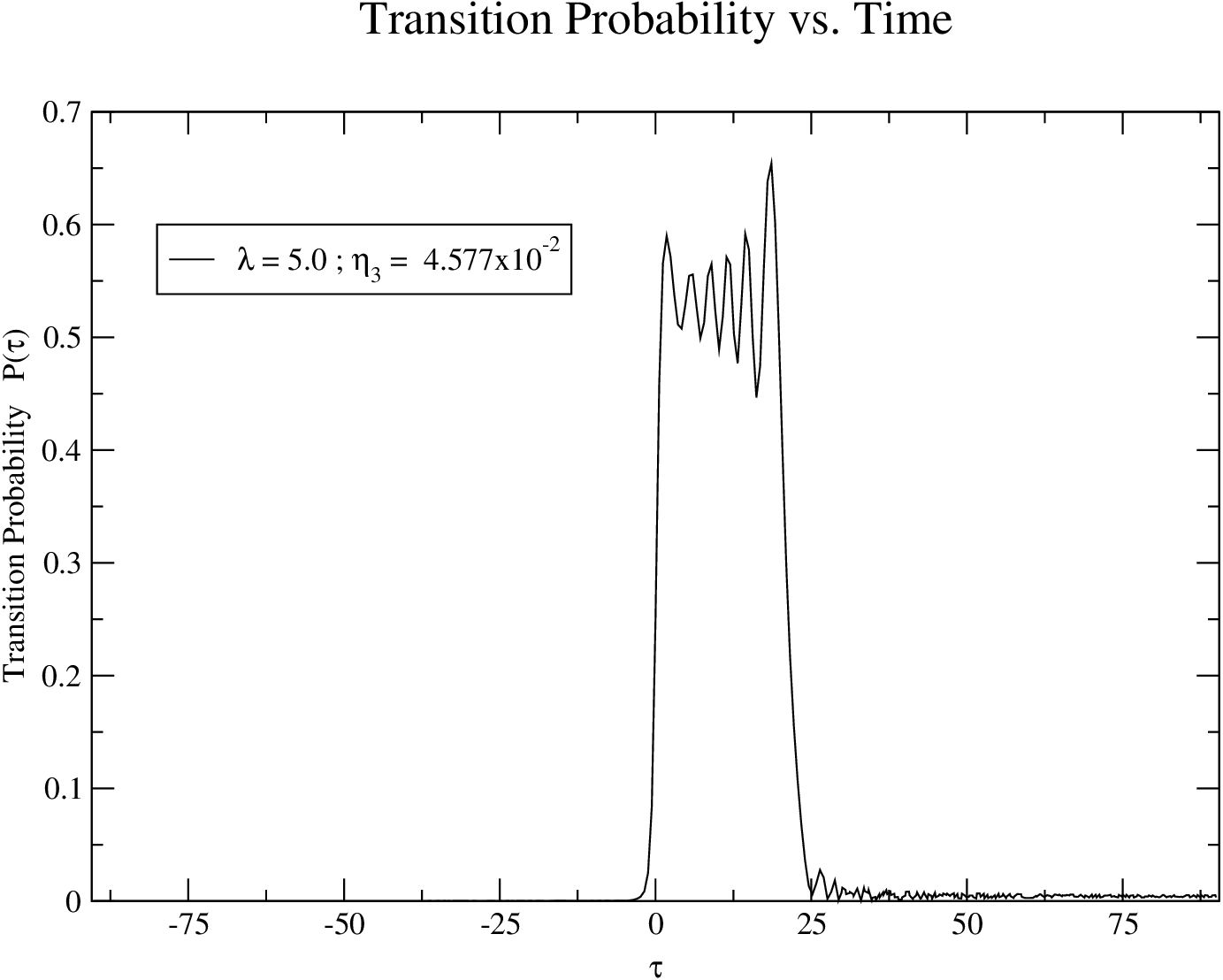}
\caption{\label{fig11} The transition probability $P(\tau )$ for non-adiabatic
rapid passage with cubic twist with $\lambda = 5.0$ and 
$\eta_{\, 3} = 4.577\times 10^{-2}$. Note the slightly reduced vertical 
scale.}
\end{figure}

We now show that one can utilize the interference between avoided crossings
to strongly suppress qubit transitions during \textit{non}-\textit{adiabatic}
rapid passage with cubic twist. Figure~\ref{fig11} shows $P(\tau )$ for
$\lambda = 5.0$ and $\eta_{\, 3} = 4.577\times 10^{-2}$.
The asymptotic transition probability for this case is 
$P = 3.44\times 10^{-3}$. This is to be compared with twistless rapid passage
with $\lambda = 5.0$ (Figure~\ref{fig4}) for which $P = 0.533$. 
Destructive interference between the two avoided crossings has reduced the
transition probability $P$ by 2 orders of magnitude relative to the 
twistless case shown in Figure~\ref{fig4}. Thus if we were to implement a 
fast NOT-operation using non-adiabatic rapid passage with cubic twist at
$\lambda = 5.0$ and $\eta_{\, 3} = 4.577\times 10^{-2}$, we would obtain 
(on average)
1 bit-flip error per 291 NOT-operations. By comparison, twistless rapid
passage with $\lambda = 5.0$ would produce (on average) 1 bit-flip error
for every 2 NOT-operations. This result strongly suggest the value of
exploring whether this destructive interference between 
avoided crossings during twisted rapid passage could be exploited to produce 
fast reliable quantum NOT and CNOT logic gates. As striking as this result for 
cubic twist is, we shall see in the following Section that quartic twist can 
reduce the bit-flip error probability even more dramatically. 

\section{\label{sec4}Quartic Twist}

For quartic twist $\phi_{4}(t) = (1/2)Bt^{4}$ and $\eta_{\, 4} = \hbar Bb^{2}
/a^{3}$. Avoided crossings are expected to occur at $\tau_{1} = 0$, and 
at $\tau_{2} = \pm 1/\sqrt{\eta_{\, 4}}$ (when $sgn\,\eta_{\, 4} = +1$;
see eqn.~(\ref{davdx2}) and Table~\ref{table1}). Formally, the analysis of 
quartic twist parallels 
that of quadratic and cubic twist. With the substitution $\phi_{2}(t)
\rightarrow \phi_{4}(t)$, eqns.~(\ref{eigstates}) and (\ref{allmisc}) continue 
to apply, and one determines $\overline{\Gamma}(\tau )$ and
$\overline{\delta}(\tau )$ from eqns.~(\ref{biggam}) and (\ref{delt}). Once
$\overline{\Gamma}(\tau )$ and $\overline{\delta}(\tau )$ are known,
eqns.~(\ref{alldeom}) can be integrated numerically subject to the initial 
condition specified in eqns.~(\ref{allinicond}). 

\subsection{Demonstration of Quantum Interference}

\begin{figure}[!h]
\includegraphics[scale=0.5]{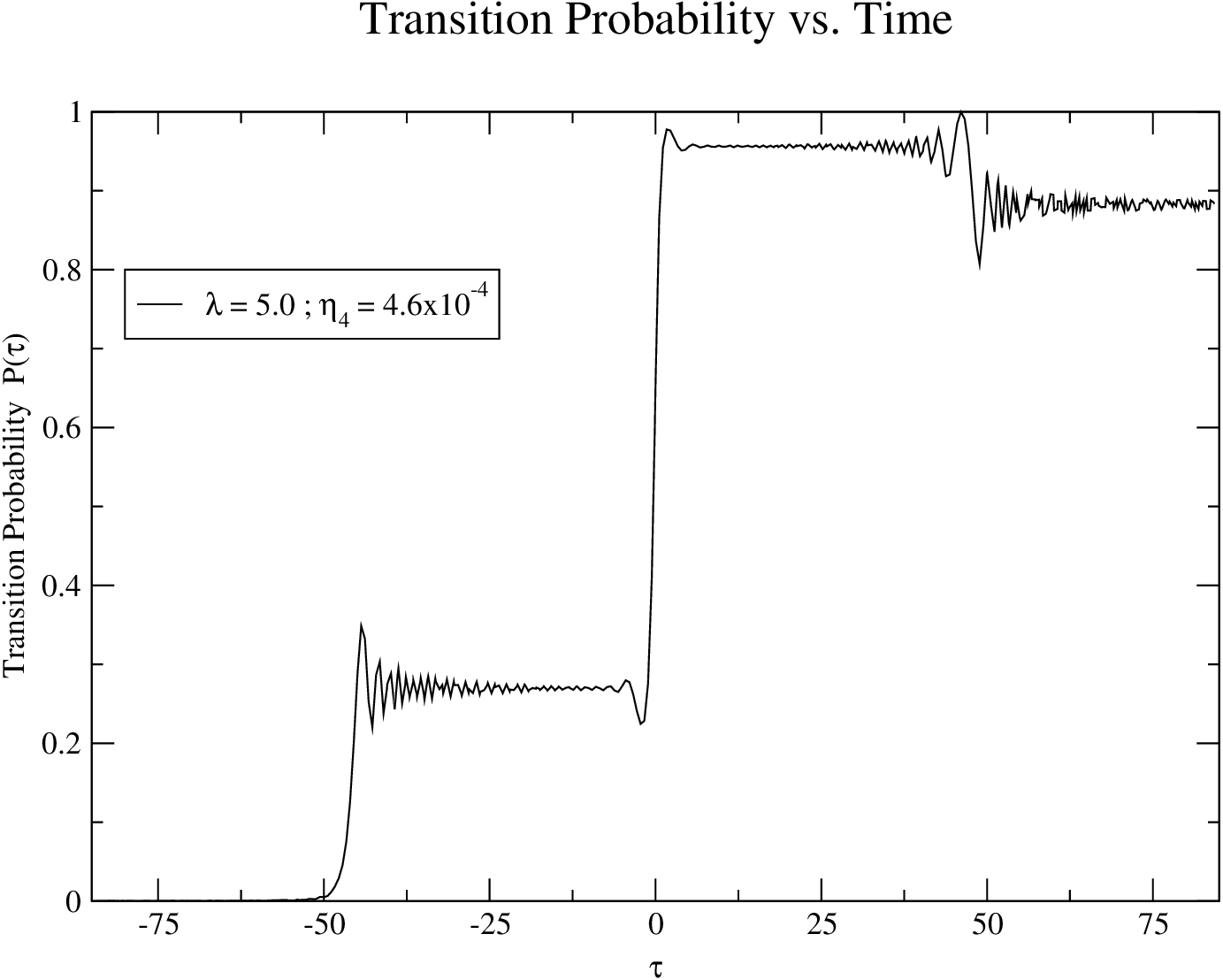}
\caption{\label{fig12} The transition probability $P(\tau )$ for
non-adiabatic rapid passage with quartic twist with $\lambda = 5.0$ and 
$\eta_{\, 4} = 4.6\times 10^{-4}$.}
\end{figure}

\begin{figure}[!h]
\includegraphics[scale=0.5]{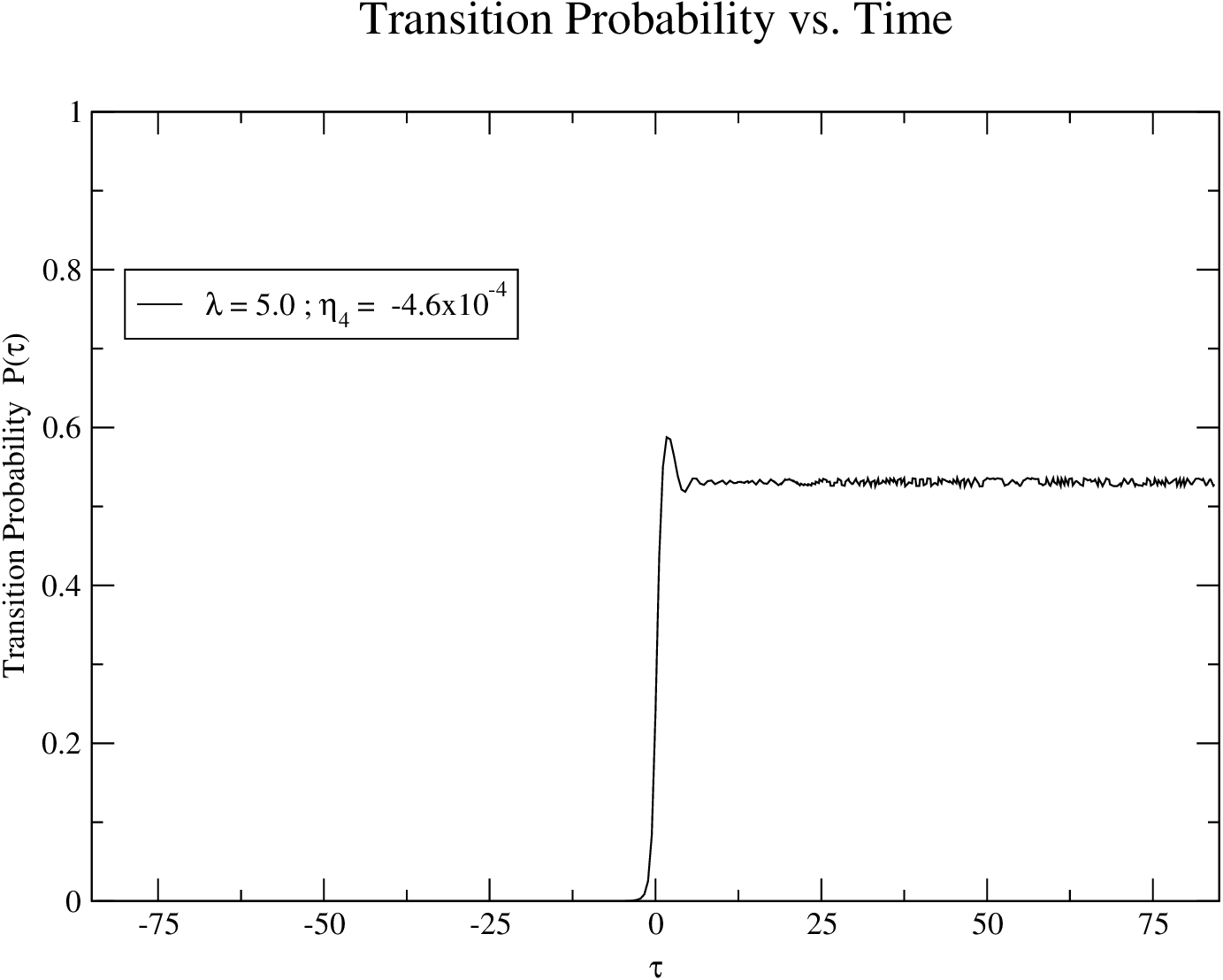}
\caption{\label{fig13} The transition probability $P(\tau )$ for
non-adiabatic rapid passage with quartic twist with $\lambda = 5.0$ and 
$\eta_{\, 4} = -4.6\times 10^{-4}$.}
\end{figure}

In Figure~\ref{fig12} we plot the transition probability $P(\tau )$ for
$\lambda = 5.0$ and $\eta_{\, 4} = 4.6\times 10^{-4}$.
The expected avoided crossings at $\tau_{1}=0$ and $\tau_{2}=\pm 46.63$
are clearly visible. The asymptotic transition probability for this case is
$P = 0.88$. For twistless rapid passage with 
$\lambda = 5.0$
(see Figure~\ref{fig4}), $P = 0.533$. Thus the avoided crossings in
Figure~\ref{fig12} are constructively interfering, leading to an enhancement
of the transition probability $P$. Figure~\ref{fig13} shows $P(\tau )$ for 
quartic twist with $\lambda = 5.0$ and $\eta_{\, 4} = -4.6\times 10^{-4}$.
This Figure clearly shows only one avoided crossing at $\tau_{1}=0$, as
expected for $sgn\,\eta_{\, 4}= -1$ (see Table~\ref{table1}). The asymptotic 
transition probability in this case is $P = 0.533$ which equals
the result for twistless rapid passage with $\lambda = 5.0$ 
(Figure~\ref{fig4}) to the level of precision obtained in our calculation.

\begin{figure}[!h]
\includegraphics[scale=0.5]{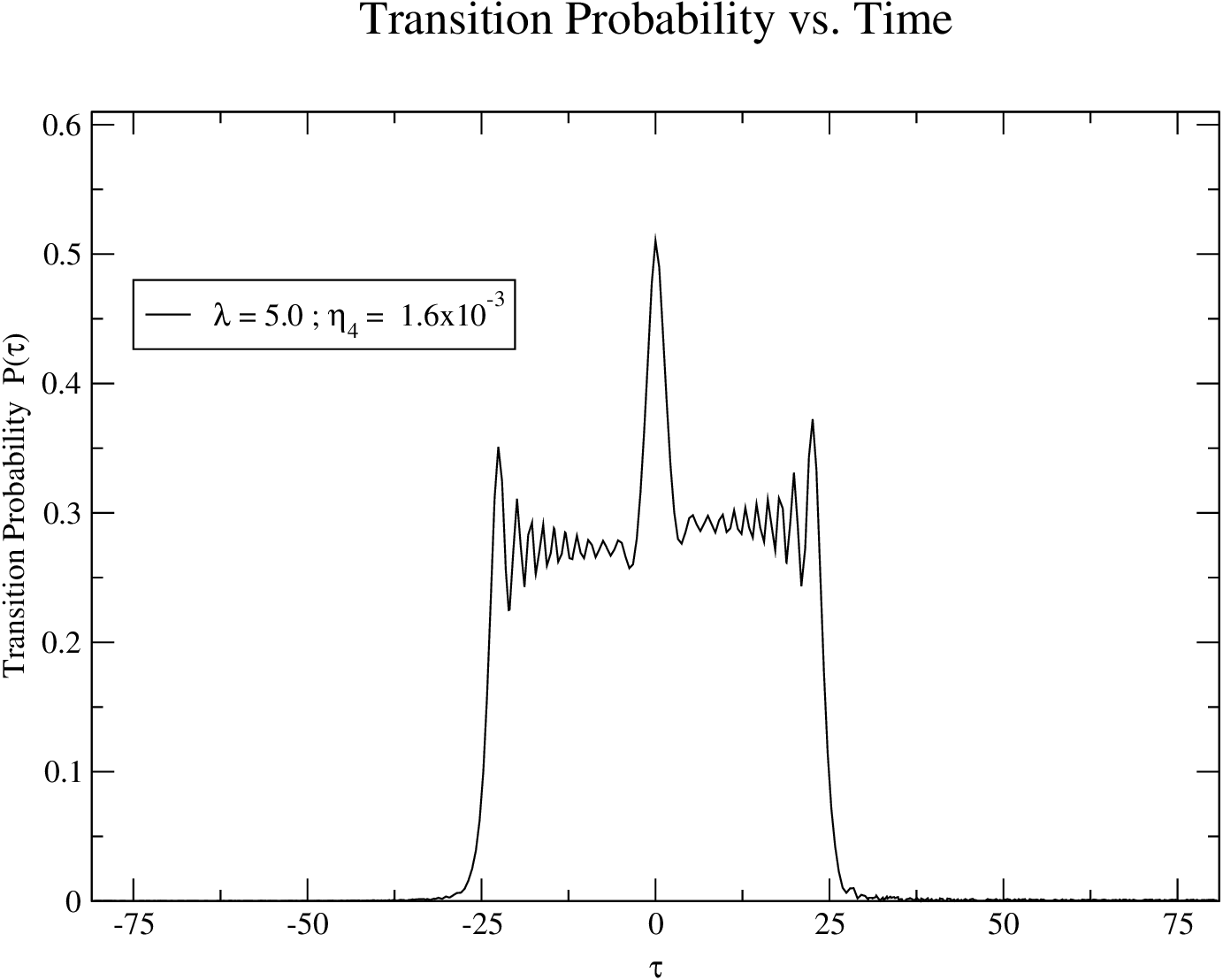}
\caption{\label{fig14} The transition probability $P(\tau )$ for
non-adiabatic rapid passage with quartic twist with $\lambda = 5.0$ and 
$\eta_{\, 4} = 1.6\times 10^{-3}$. Note the slightly reduced vertical scale.}
\end{figure}

Figure~\ref{fig14} plots $P(\tau )$ for $\lambda = 5.0$ and $\eta_{\, 4} =
1.6\times 10^{-3}$. 
The Figure clearly shows the expected crossings at $\tau_{1}=0$ and
$\tau_{2} = \pm 25.0$. The asymptotic transition probability is
$P = 6.93\times 10^{-4}$ and corresponds to destructive interference
relative to twistless rapid passage with $\lambda = 5.0$ (Figure~\ref{fig4}).
We do not include a plot of $P(\tau )$ for $\lambda = 5.0$ and $\eta_{\, 4} =
-1.6\times 10^{-3}$ as it is similar to Figure~\ref{fig13}: one avoided
crossing at $\tau_{1}=0$ and $P = 0.533$.

Summarizing these results, we see that: (i) three (one) avoided crossings 
(crossing) occur(s)
as predicted in Table~\ref{table1} when $sgn\,\eta_{\, 4} = +1\; (-1)$;
(ii) the avoided crossings produce interference effects in the
transition probability, with the character of the interference (constructive
or destructive) determined by the separation of the avoided crossings; and
(iii) the separation of adjacent avoided crossings is given by $\Delta\tau_{ac}
= |\tau_{2}-\tau_{1}| = 1/\sqrt{\eta_{\, 4}}$ ($sgn\,\eta_{\, 4}=+1$), and 
it is controllable through variation of $\eta_{\, 4}=\hbar Bb^{2}/a^{3}$.

\subsection{Non-Resonant Pump}

\begin{figure}[!h]
\includegraphics[scale=0.5]{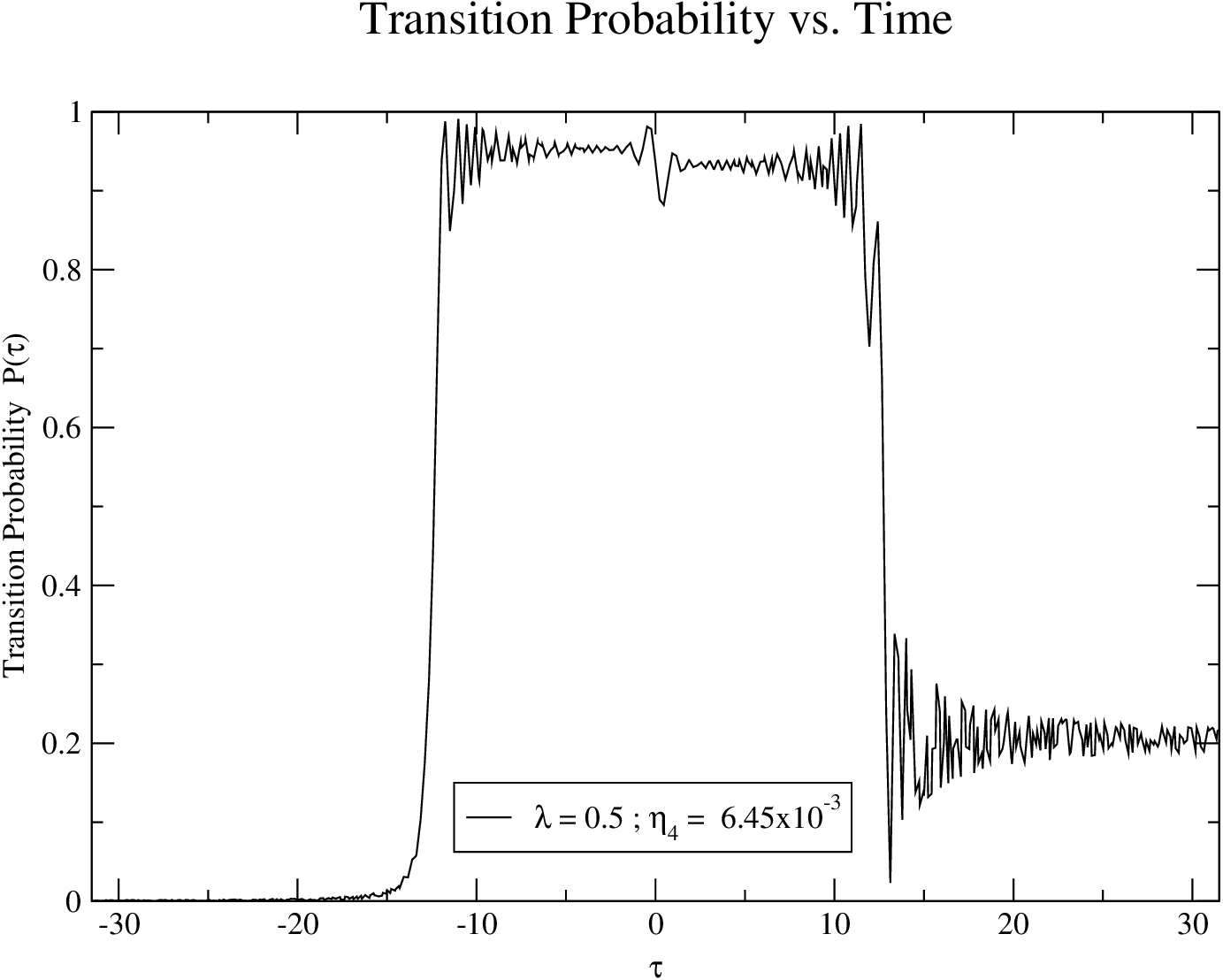}
\caption{\label{fig15} The transition probability $P(\tau )$ for
adiabatic rapid passage with quartic twist with $\lambda = 0.5$ and 
$\eta_{\, 4} = 6.45\times 10^{-3}$.}
\end{figure}

Quartic twist does not appear to be as effective at pumping the qubit 
energy-levels as cubic twist. Figure~\ref{fig15} shows $P(\tau )$ for
$\lambda = 0.5$ and $\eta_{\, 4} = 6.45\times 10^{-3}$.
The expected avoided crossings at $\tau_{1}=0$ and $\tau_{2}=\pm 12.45$
are clearly visible, and the asymptotic transition probability is
$P = 0.20$. Although this is a 2 order of magnitude improvement over
twistless adiabatic rapid passage with $\lambda = 0.5$ (Figure~\ref{fig9}),
it falls well short of the transition probability $P=0.996$ easily obtainable
with cubic twist. In fact, for $\eta_{\, 4}<1$, $P\sim 0.20$ was among the
largest $P$-values we could find. If larger values of twist strength are
allowed, the largest transition probability we could find was $P=0.64$
at $\eta_{\, 4}=3.00$.

\subsection{\label{sec4c}Transition Quenching}

Quartic twist proves to be much more effective at quenching 
transitions during non-adiabatic rapid passage than cubic twist. 
Table~\ref{table2} gives the transition probabilities for quartic twist
pulses for which $\lambda = 5.00$ and $\eta_{4}$ lies in the interval
[$3.95\times 10^{-3}$, $4.04\times 10^{-3}$]. 
\begin{table}[!h]
\caption{\label{table2}Transition probabilities for quartic twist with
$\protect\lambda = 5.00$ and $\protect\eta_{4}$ in the range [$3.95\times 
10^{-3}$, $4.04\times 10^{-3}$]\vspace{0.3in}.} 
\begin{ruledtabular}
\begin{tabular}{cc}
$\:\eta_{4} \;\; (\,\times 10^{-3}\, )\; $ & \, $P$ \\ \hline
3.95  & $2.0\times 10^{-2} $ \\
3.96  & $1.3\times 10^{-2} $ \\
3.97  & $6.8\times 10^{-3} $ \\
3.98  & $3.6\times 10^{-3} $ \\
3.99  & $9\times 10^{-4} $ \\ 
4.00  & $4\times 10^{-5} $ \\
4.01  & $8\times 10^{-4} $ \\
4.02  & $3.9\times 10^{-3} $ \\
4.03  & $1.0\times 10^{-2} $ \\
4.04  & $1.7\times 10^{-2} $ \\ 
\end{tabular}
\end{ruledtabular}
\end{table}
The essential thing to notice in Table~\ref{table2} is that for $\eta_{4}
= 4.00\times 10^{-3}$, the transition probability $P = 4\times 10^{-5}$.
This is significant for the following reason. It has been shown that a quantum
computation of arbitrarily long duration becomes possible if the quantum
logic gates used to implement the computation all have error probabilities
(per gate operation) which lie below the threshold $P_{ft}$ for fault tolerant
operation \cite{pre}. This threshold has been estimated to be $P_{ft} \sim
10^{-4}-10^{-5}$ \cite{flt}. In terms of the gate fidelity $F = 1 - P$,
the more optimistic estimate for $P_{ft}$ gives $F_{ft} = 0.9999$. We see 
that for $\lambda = 5.00$ and $\eta_{4} = 4.00\times 10^{-3}$, twisted rapid 
passage with quartic twist gives a gate fidelity of $F = 0.99996$ which 
exceeds the best case estimate for fault tolerant operation $F_{ft} = 0.9999$. 
This fault tolerant performance is achieved while inverting the qubit at a 
\textit{non}-\textit{adiabatic} rate. 
The reader should note that the values $\lambda = 5.00$ and $\eta_{4} =
4.0\times 10^{-3}$ can be realized with existing NMR technology (see 
Section~\ref{sec5d}). Our analysis
raises the exciting possibility that non-adiabatic rapid passage with quartic 
twist might provide a means of realizing \textit{fast fault-tolerant} NOT and 
CNOT gates. The novelty of this prospect is the marriage of operational speed 
with fault-tolerance. 
This marriage of speed and reliability is a direct consequence of the 
destructive interference which is possible between the 3 avoided crossings 
that arise during rapid passage with quartic twist. Quantum CNOT gates
are ubiquitous in quantum computing and quantum error correction
\cite{uni,bar,qec}. Thus, determining how to implement them in a fast
fault tolerant manner is a potentially significant development for the field.

\section{\label{sec5}Discussion}

\subsection{Summary}

It has been our aim in this paper to show that multiple avoided crossings can
arise during twisted rapid passage, and that by varying their time-separation,
interference effects are produced which allow for a direct control over
qubit transitions. This time-separation is controlled through the
(dimensionless) twist strength $\eta$, and the resulting interference can be
constructive (enhancing transitions) or destructive (reducing 
transitions). For nth-order polynomial twist, $\eta_{\, n} = \hbar Bb^{n-2}/
a^{n-1}$, where $B$ is the (dimensionful) twist strength, $2b$ is the 
energy-gap separating the qubit energy-levels at an avoided crossing, and $a$ 
is the inversion rate of the external field $\mathbf{F}(t)$ (see 
Section~\ref{sec2}). The interference effects are a consequence of the
temporal phase coherence of the wavefunction. The ability to vary this
coherence by varying the temporal separation of the avoided crossings renders
twisted rapid passge with adjustable twist strength into a temporal
interferometer through which qubit transitions can be greatly enhanced or
suppressed. Cubic and quartic twist were explicitly considered in 
this paper as they are, respectively, the simplest examples of odd-order and 
even-order polynomial twist in which these interference effects are 
expected to occur. Although we have focused on these two cases, we do not
mean to suggest that these pulses are the best of all possible twisted
rapid passage pulses. A search is currently underway for other twisted rapid
passage pulses that might produce stronger destructive interference, and hence,
faster, more fault tolerant quantum CNOT and NOT gates (see below for
further discussion). We have seen that this interference mechanism 
can be used to pump qubit energy-levels, as well as to strongly quench 
qubit transitions during \textit{non}-\textit{adiabatic} twisted rapid 
passage. Although cubic twist proved to be more effective at pumping than 
quartic twist, quartic twist was found to be much more effective at quenching 
qubit transitions. We have seen that quartic twist allows qubit inversion
to be done both non-adiabatically and at fidelities that exceed the threshold
for fault tolerant operation. The marriage of operational speed with 
reliability is a direct consequence of the destructive interference that is 
possible between the 3 avoided crossings that can arise during rapid passage 
with quartic twist.

\subsection{Implementing Quantum CNOT Gate}

\begin{figure}[h]
\includegraphics[scale=0.5]{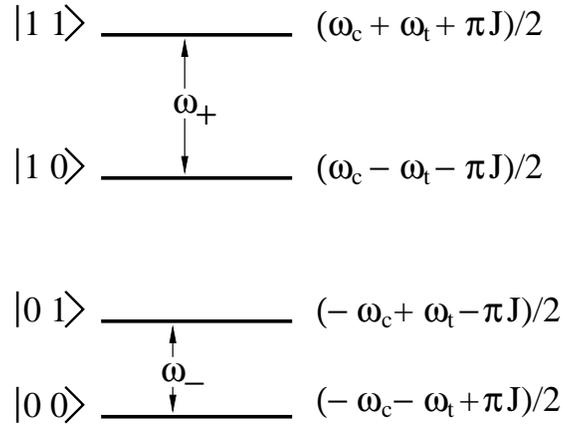}
\caption{\label{fig16} Energy-level structure appropriate for implementing a 
quantum CNOT operation using twisted rapid passage. The corresponding energies 
(in frequency units) appear to the right of the energy levels.}
\end{figure}

We now describe a procedure for implementing a quantum CNOT gate using
twisted rapid passage in the context of liquid state NMR. If the
liquid has low viscosity, one can ignore dipolar coupling between the
qubits, and if the remaining Heisenberg interaction between the qubits is
weak compared to the individual qubit Zeeman energies, it can be 
well-approximated by an Ising interaction \cite{chu}. Under these
conditions, the Hamiltonian (in frequency units) for the control ($c$) and 
target ($t$) qubits is:
\begin{equation}
\frac{H_{ct}}{\hbar} = -\omega_{c}\, I^{c}_{z} -\omega_{t}\, I^{t}_{z}
                        + 2\pi J\, I^{c}_{z}\,I^{t}_{z} \hspace{0.1in} .
\end{equation}
Here $\omega_{c}$ ($\omega_{t}$) is the resonance frequency of the isolated
control (target) qubit; $J$ is the Ising coupling constant; and
$\omega_{c} > \omega_{t} > \pi J$. We choose the single qubit computational
basis states (CBS) to be $|0\rangle = |\uparrow\rangle$ and $|1\rangle =
|\downarrow\rangle$. Thus the two-qubit CBS are: $|00\rangle = |\uparrow
\uparrow\rangle$; $|01\rangle = |\uparrow\downarrow\rangle$; $|10\rangle =
|\downarrow\uparrow\rangle$; and $|11\rangle = |\downarrow\downarrow\rangle$,
and they are the eigenstates of $H_{ct}$. The energy-levels (in frequency 
units) are shown in Figure~\ref{fig16}, where
\begin{equation}
\omega_{\pm} = \omega_{t} \pm \pi J \hspace{0.1in} .
\end{equation}
\noindent Given this energy-level structure, we can implement a quantum CNOT 
operation on the two qubits by sweeping through the $\omega_{+}$ resonance 
using
twisted rapid passage. Decoupling \cite{ch2} is used to switch off the 
dynamics of the
control qubit so that only the target qubit responds to the rapid passage 
pulse. Since the two states $|00\rangle$ and $|01\rangle$ are not resonant, 
they do not respond to the twisted rapid passage pulse. Thus,
\begin{equation}
\begin{array}{l}
|00\rangle\rightarrow |00\rangle \\
|01\rangle\rightarrow |01\rangle \hspace{0.1in} .
\end{array}
\end{equation}
On the other hand, for the $|10\rangle$ and $|11\rangle$ states, the 
combination of decoupling and sweeping through the $\omega_{+}$ resonance 
means that only the target qubit has its spin flipped. Thus,
\begin{equation}
\begin{array}{l}
|10\rangle\rightarrow |11\rangle \\
|11\rangle\rightarrow |10\rangle \hspace{0.1in} ,
\end{array}
\end{equation}
and we see that this procedure implements a quantum CNOT operation on the 
two qubits.

\subsection{Experimental Realization}

Because of the fundamental significance of quantum CNOT gates to quantum
computing and quantum error correction \cite{uni,bar,qec}, it is hoped that
the feasibility of using rapid passage with quartic twist to implement this
gate might be tested experimentally (see penultimate paragraph of 
Section~\ref{sec1}). Experimental realization of polynomial 
twist $\phi_{n}(t) = (2/n)Bt^{n}$ should be possible through an adaptation of 
the procedure used by Zwanziger et.\ al.\ \cite{zwa} to realize quadratic 
twist. Thus: (1) the driving rf-field
is linearly polarized along the x-axis in the lab-frame with $F_{x}(t) 
= 2b\cos\phi_{rf}(t)$; (2) the resonance offset $at$ (see eqn.~(\ref{labham}))
is produced by linearly sweeping the detector frequency $\omega_{det}(t)$ 
through the resonance at the Larmor frequency $\omega_{0}$ such that 
$\omega_{det}(t) = \omega_{0} + (2at/\hbar )$; and (3) twist is introduced by 
sweeping the rf-frequency $\omega_{rf}(t) = \dot{\phi}_{rf}$ through the 
resonance at $\omega_{0}$ in such a way that $\omega_{rf}(t) = \omega_{det} - 
\dot{\phi}_{n}$. It is worth noting that the resonance condition 
$\omega_{rf}(t) =\omega_{0}$ is identical to our existence condition 
for avoided crossings, eqn.~(\ref{cond}). Note that in our paper
the external field inversion takes place over the time-interval ($-T_{0}/2$,
$T_{0}/2$); the external field crosses the x-y plane at $t=0$ and is 
initially aligned along the $-\hat{\mathbf{z}}$ direction. The Appendix
provides a translation key which relates the theoretical parameters of this 
paper
to the experimental parameters of the Zwanziger experiments \cite{jwz,zwa}.

Before leaving the subject of experimental realization of rapid passage
with quartic twist, two further remarks are in order. First, to insure
that all qubits are inverted when a spread of resonance frequencies occurs,
it is necessary to require that the frequency sweep cover a large enough
interval that the entire spread of resonance frequencies is included in it. 
This gaurantees that all qubits will have passed through resonance by the end 
of the frequency sweep. Second, a range of rf field strengths can also be 
accomodated so long as $aT/2 \gg b_{max}$. This condition insures that the 
frequency sweep begins far from resonance for all rf field strengths, and that 
transitions will continue to occur only near the avoided crossings. One 
therefore anticipates that in this case also, the interference effects will 
continue to occur as predicted. 
For reasonably good samples, magnets, and rf sources these constraints can be 
satisfied, and the interference effects presented above should be readily
observable. This is in fact what is found experimentally~\cite{jwz}.

\subsection{\label{sec5d}Other Pulses}

Having introduced twisted rapid passage with polynomial twist, and pointed out
the possible advantages of quartic twist for quantum computing, it is natural
to ask how quartic twist compares with the more familiar $\pi$-pulse which
can also be used to implement quantum CNOT and NOT gates. We begin by comparing
the inversion time for quartic twist with that of a comparable $\pi$-pulse.
We focus on quartic twist with $\lambda = 5.00$ and $\eta_{4}= 4.00\times
10^{-3}$ as this choice of parameters yields a gate fidelity $F=0.99996$
(see Section~\ref{sec4c}) which exceeds the threshold for fault tolerant
gate operation $F_{ft}\sim 0.9999$. We now show that this case achieves fault 
tolerant operation while simultaneously matching the inversion speed of a 
$\pi$-pulse. 
In the notation of Ref.~\cite{zwa}, the basic experimental parameters for 
twisted rapid passage are $A$, $B$, $\omega_{1}$, and $T$, and they are 
related to our theoretical parameters by eqns.~(\ref{A8}) and (\ref{A3}).
$T$ continues to denote the duration of the twisted rapid passage pulse. 
Because a twisted rapid passage sweep must begin 
far from the avoided crossings, $A$ and $\omega_{1}$ cannot be chosen 
independently. In the rf-frame, the asymptotic effective magnetic field must 
lie near the z-axis so that
$\tan\theta = \omega_{1}/A \sim 0.1$. Choosing $\omega_{1} = 4000$ Hz gives
$A = 4\times 10^{4}$ Hz. Both of these values can be achieved with existing 
NMR technology. Writing $f = \omega_{1}/A$, eqn.~(\ref{A9}) gives
\begin{displaymath}
T_{4} = \frac{4}{f\omega_{1}\lambda} \hspace{0.1in} .
\end{displaymath}
With $\lambda = 5.00$, this gives
\begin{displaymath}
T_{4} = 2\; \textrm{msec} \hspace{0.1in} .
\end{displaymath}
By comparison, the inversion time for a $\pi$-pulse with rf-amplitude
$\omega_{1} = 4000$ Hz is $T_{\pi} = \pi /\omega_{1} = 0.8$ msec. Thus,
twisted rapid passage with quartic twist is clearly capable of matching the 
inversion speed of a comparable $\pi$-pulse while still exceeding the 
threshold for fault tolerant operation. On the other hand, the error 
probability for a typical $\pi$-pulse is $P\sim 10^{-3}$ due to, 
for example, inhomogenities in the rf field amplitude. This corresponds to a 
fidelity $F \sim 0.999$ so that, unlike the equally fast quartic twist pulse 
which acts
fault tolerantly, the $\pi$-pulse falls \textit{short\/} of the threshold for 
fault tolerant operation $F_{th} \sim 0.9999$.

We hope in the future to examine higher order versions of polynomial twist
to determine whether they have more effective quenching and/or robustness
properties than cubic and quartic twist. We have also done preliminary work on 
the interesting case of periodic twist: $\phi (t) = \pi\rho\sin\omega t$. 
As we have seen, polynomial twist only allows 1--3 avoided crossings to occur 
during rapid passage. One can show that periodic twist allows the number of 
avoided crossings that occur during rapid passage to be modified through 
variation of the twist amplitude $\rho$ and frequency $\omega$. We intend 
to explore how the interference effects considered here are modified when more 
than 3 avoided crossings can occur.

\begin{acknowledgments}
I would like to thank: (1) T. Howell III for continued support; and 
(2) the National Science Foundation for support provided through 
grant number NSF-PHY-0112335. 
\end{acknowledgments}

\appendix
\section{Connection Between Theory and Experiment}

For ease of comparison with Refs.~\cite{zwa} and \cite{jwz}, we choose
$\mathbf{F}(t) = -b\cos\phi_{n}(t)\hat{\mathbf{x}} -b\sin\phi_{n}(t)
\hat{\mathbf{y}} + at\hat{\mathbf{z}}$ in eqn.~(\ref{labham}). The
Hamiltonian in the detector frame is then
\begin{displaymath}
\frac{H(t)}{\hbar} = \frac{at}{\hbar}\,\sigma_{z} - \frac{b}{\hbar}\,
                      \cos\phi_{n}(t)\,\sigma_{x} - 
                        \frac{b}{\hbar}\,\sin\phi_{n}(t)\,
                           \sigma_{y} \hspace{0.1in} .
\end{displaymath} 
Here $\phi_{n}(t) = (2/n)\mathcal{B}t^{n}$, and to avoid confusion with the
notation of Ref.~\cite{zwa}, we have switched the symbol used for the
twist strength in the main body of this paper: $B\rightarrow \mathcal{B}$.
Transformation to the rf-frame is done using the unitary operator $U(t) =
\exp [-(i/2)\,\phi_{n}(t)\,\sigma_{z}]$ so that $H(t)\rightarrow 
\overline{H}(t)$:
\begin{eqnarray}
\frac{\overline{H}(t)}{\hbar} & = & \left(\,\frac{at}{\hbar} -
                                      \frac{\dot{\phi}_{n}}{2}\,\right)\,
                                       \sigma_{z} - \frac{b}{\hbar}\,
                                        \sigma_{x} \nonumber \\
 & = & \left(\,\frac{2at}{\hbar} - \dot{\phi}_{n}\,\right)\, I_{z} -
        \frac{2b}{\hbar}\, I_{x} \hspace{0.1in} , \label{A1}
\end{eqnarray}
and $\mathbf{I} = \mathbf{\sigma}/2$.

The experimental Hamiltonian in the rf-frame appears in eqn.~(12) of 
Ref.~\cite{zwa}:
\begin{equation}
\frac{\overline{H}_{ex}(t)}{\hbar} = \left(\,\dot{\phi}_{rf} - \omega_{0}\,
                                       \right)\, I_{z} - \omega_{1}\, I_{x}
   \hspace{0.1in} . \label{A2}
\end{equation}
Comparing eqns.~(\ref{A1}) and (\ref{A2}) gives
\begin{equation}
\omega_{1} = \frac{2b}{\hbar}
 \label{A3}
\end{equation}
and
\begin{equation}
\dot{\phi}_{rf} - \omega_{0} = \frac{2at}{\hbar} - \dot{\phi}_{n}
  \hspace{0.1in} . \label{A4}
\end{equation}
Integrating eqn.~(\ref{A4}) gives
\begin{equation}
\phi_{rf}(t^{\prime\prime}) = \int_{-T/2}^{t^{\prime\prime}}\, 
                                dt^{\prime\prime\prime}\,\left[\,\omega_{0} +
                              \frac{2aT}{\hbar}\left(\,
                               \frac{t^{\prime\prime\prime}}{T}\,\right) -
                        2\mathcal{B}T^{n-1}\,\left(\,
                         \frac{t^{\prime\prime\prime}}{T}\,\right)^{n-1}
                          \,\right] \hspace{0.1in} . \label{A5}
\end{equation}
In this paper, we have parameterized time such that $t^{\prime\prime\prime}
\in  [\, -T/2, T/2\,]$, and $T$ is the duration of the twisted rapid
passage pulse. Defining
\begin{displaymath}
\tau = \frac{t^{\prime\prime\prime}}{T} + \frac{1}{2} \hspace{0.1in} ,
\end{displaymath}
it follows that $\tau\in [0, 1]$. Introducing $t^{\prime} =
t^{\prime\prime\prime}+ T/2$ and $t = t^{\prime\prime} + T/2$, eqn.~(\ref{A5})
becomes
\begin{equation}
\phi_{rf}(t) = \int_{0}^{t}\, dt^{\prime}\,\left[\,\omega_{0} +
                \left(\, \frac{2aT}{\hbar}\,\right)\,\left(\,\tau -
                 \frac{1}{2}\,\right) - \frac{2\mathcal{B}T^{n}}{T}\,
                  \left(\,\tau - \frac{1}{2}\,\right)^{n-1}\,\right]
                   \hspace{0.1in} . \label{A6}
\end{equation}

As explained in the caption of Figure~2 of Ref.~\cite{zwa}, 
$\dot{\phi}_{rf} = \dot{\phi}_{det} - \dot{\phi}_{n}$; with $\dot{\phi}_{det}
= \omega_{0} + 2A(\tau - 1/2)$; and generalizing to polynomial twist,
$\dot{\phi}_{n} = nB(\tau - 1/2)^{n-1}/T$, where $B$ is the symbol used in
Ref.~\cite{zwa} for the twist strength. Plugging these expressions for
$\dot{\phi}_{det}$ and $\dot{\phi}_{n}$ into $\dot{\phi}_{rf} =
\dot{\phi}_{det} - \dot{\phi}_{n}$, and integrating gives
\begin{equation}
\phi_{rf}(t) = \int_{0}^{t}\, dt^{\prime}\,\left[\,\omega_{0} +
                      2A\left(\,\tau - \frac{1}{2}\,\right) - \frac{nB}{T}\,
                       \left(\,\tau - \frac{1}{2}\,\right)^{n-1}\,\right]
                        \hspace{0.1in} . \label{A7}
\end{equation}
Equating eqns.~(\ref{A6}) and (\ref{A7}) gives
\begin{subequations}
 \label{A8}
  \begin{eqnarray}
     A & = & \frac{aT}{\hbar}  \label{A8a} \\
     B & = & \frac{2\mathcal{B}}{n}T^{n} \hspace{0.1in} . \label{A8b}
  \end{eqnarray}
\end{subequations}
Using eqns.~(\ref{A3}) and (\ref{A8}) in the definition of $\lambda$ (see
discussion following eqns.~(\ref{allmisc})) gives 
\begin{equation}
\lambda = \frac{4\, |A|}{\omega_{1}^{2}T} \hspace{0.1in} . \label{A9}
\end{equation}
Using eqns.~(\ref{A3}), (\ref{A8}) and eqn.~(\ref{twststr}) with $n=3$ and $4$ 
gives
\begin{equation}
\eta_{3} = \frac{3}{4}\,\frac{B\omega_{1}}{A^{2}T} \label{A10}
\end{equation}
and
\begin{equation}
\eta_{4} = \frac{B\omega_{1}^{2}}{2A^{3}T} \hspace{0.1in} ,
\label{A11}
\end{equation}
respectively. The results of this Appendix give the connection between our
theoretical parameters and the experimental
parameters $A$, $B$, $\omega_{1}$, and $T$ of the Zwanziger experiments
\cite{jwz,zwa}. In Section~\ref{sec5}, these formulas are used to calculate 
the inversion time for a twisted rapid passage pulse with quartic twist.

\end{document}